\documentclass{aa}

\bibpunct{(}{)}{;}{a}{}{,} 
\usepackage{natbib}
\usepackage{graphicx}
\usepackage{rotating}

\usepackage[varg]{txfonts}
\begin{document} 
   \title{A re-analysis of the NuSTAR and XMM-Newton broad-band spectrum of Ser~X-1}
%
%
  \author{M. Matranga\inst{1}
          \and
          T. Di Salvo\inst{1}
          \and
          R. Iaria\inst{1}
          \and
          A. F. Gambino\inst{1}
          \and
          L. Burderi\inst{2}
         \and
          A. Riggio\inst{2}
          \and
          A. Sanna\inst{2}
}
   \institute{Universit\'a degli Studi di Palermo, Dipartimento di Fisica e Chimica,  
via Archirafi 36, 90123 Palermo, Italy\\
             \email{marco.matranga01@unipa.it}
         \and
             Universit\'a degli Studi di Cagliari, Dipartimento di Fisica, SP 
Monserrato-Sestu KM 0.7, 09042 Monserrato, Italy\\
            }


 %
   \abstract
   {High resolution X-ray spectra of neutron star Low Mass X-ray Binaries 
(LMXBs) in the energy range 6.4-6.97 keV,
are often characterized by the presence of K$\alpha$ transition features of iron 
at different ionization stages. Since these lines are thought to originate by  
reflection of the primary Comptonization spectrum over the accretion disk, 
the study of these features allows us to investigate the structure of the 
accretion flow close to the central source.
Thus, the study of these features gives us important physical information 
on the system parameters and geometry.
Ser X-1 is a well studied LMXB which clearly shows a broad iron line.
Several attempts to fit this feature as a smeared reflection feature have
been performed on \emph{XMM-Newton}, \emph{Suzaku}, \emph{NuSTAR}, 
and, more recently, on  \emph{Chandra} data, 
finding different results for the inner radius of the disk
and other reflection or smearing parameters. 
For instance, \citet{Miller.etal:13} have presented broad-band, high quality 
\emph{NuSTAR} data of Ser~X-1. 
Using relativistically smeared self-consistent reflection models, 
they find a value of R$_{in}$ close to 1.0 R$_{ISCO}$ (corresponding to 6 R$_g$,
where R$_g$ is the Gravitational radius, defined as usual R$_g = G M / c^2$), and 
a low inclination angle of less than $\sim 10^\circ$.}
   {The aim of this paper is to probe to what extent the choice of reflection 
and continuum models (and uncertainties therein) can affect the conclusions about the 
disk parameters inferred from the reflection component. To this aim we re-analyze all 
the available public NuSTAR and XMM-Newton which have the best sensitivity at the 
iron line energy observations of Ser X-1. Ser X-1 is a well studied source, its
spectrum has been observed by several instruments, and is therefore one of the best 
sources for this study.}
   {We use slightly different continuum and reflection models with
respect to those adopted in literature for this source. In particular
we fit the iron line and other reflection features with self-consistent
reflection models as reflionx  (with a power-law illuminating continuum 
modified with a high energy cutoff to mimic the shape of the incident Comptonization
spectrum) and rfxconv. With these models we fit
\emph{NuSTAR and \emph{XMM-Newton} spectra yielding consistent spectral results.}}
   {Our results are in line with those already found by \citet{Miller.etal:13} 
but less extreme. In particular, we find the inner disk radius at $\sim 13 \, R_g$
and an inclination angle with respect to the line of sight of $\sim 27^\circ$. 
We conclude that, while the choice of the reflection model has little impact
on the disk parameters, as soon as a self-consistent model is used, the choice of 
the continuum model can be important in the precise determination of the disk 
parameters from the reflection component. Hence broad-band X-ray spectra are highly 
preferable to constrain the continuum and disk parameters.}
   {}


   \maketitle

\section{Introduction}
\label{sec:intro}
X-ray spectra emitted by Low Mass X-Ray Binaries (LMXBs) of the atoll 
class \citep{Hasinger.etal:89} are usually characterized by two states 
of emission: the soft and the hard state. 
During soft states the spectrum can be well described by a 
soft thermal component, usually a blackbody or a disk multi-color 
blackbody, possibly originated from the accretion disk, and a harder
component, usually a saturated Comptonization spectrum. In some cases, 
a hard power-law tail has been detected in the spectra of these sources
during soft states both in Z sources \citep{DiSalvo.et:00}, and in atoll 
sources \citep[e.g.,][]{Piraino.etal:07}, usually interpreted as 
Comptonization off a non-thermal population of electrons. On the other 
hand, during hard states the hard component of the spectrum can be 
described by a power law with high energy cutoff, interpreted as unsaturated 
Comptonization, and a weaker soft blackbody component 
\citep[e.g.,][]{DiSalvo.etal:15}. The hard component 
is generally explained in terms of inverse Compton scattering of soft 
photons, coming from the neutron star surface and/or the inner accretion 
disk, by hot electrons present in a corona possibly located in the inner
part of the system, surrounding the compact object \citep{DAi.etal:10}. 

In addition to the continuum, broad emission lines in the range 6.4-6.97 
keV are often observed in the spectra of LMXBs \citep[see e.g.][]{Cackett.etal:08, 
Pandel.etat:08, DAi.etal:09, DAi.etal:10, Iaria.etal:09, DiSalvo.etal:05, 
DiSalvo.etal:09, Egron.etal:13, DiSalvo.etal:15}. These lines are 
identified as K$\alpha$ transitions of iron at different ionization 
states and are thought to originate from reflection of the primary 
Comptonization spectrum over the accretion disk. 
These features are powerful tools to investigate the structure of the 
accretion flow close to the central source. In particular, important 
information can be inferred from the line width and profile, since the 
detailed profile shape is determined by the ionization state, geometry 
and velocity field of the emitting plasma \citep[see e.g.][]{Fabian.etal:89}.
Indeed, when the primary Comptonization spectrum illuminates a colder 
accretion disk, other low-energy discrete features (such as emission lines
and absorption edges) are expected to be created by photoionization and 
successive recombination of abundant elements in different ionizations states 
as well as a continuum emission caused by direct Compton scattering of the
primary spectrum off the accretion disk. All these features together 
form the so-called reflection spectrum, and the whole reflection spectrum 
is smeared by the velocity-field of the matter in the accretion disk.

Ser~X-1 is a persistent accreting LMXB classified as an atoll source, 
that shows type~I X-ray bursts. 
The source  was discovered in 1965 by
\citet{Friedman.etal:67}. \citet{Li.etal:76} firstly discovered type-I X-ray bursts 
from this source that was therefore identified as an accreting neutron star.
Besides type-I bursts with typical duration of few 
seconds \citep{Balucinska.etal:85}, a super-burst of the
duration of about 2 hours has also been reported \citep{Cornelisse.etal:02}.
Recently \citet{Cornelisse.etal:13}, analyzing spectra collected by GTC, 
detected a two-hours periodicity. They tentatively identified this periodicity 
as the orbital period of the binary and hence proposed that the secondary star  
might be a main sequence K-dwarf.


\cite{Church.etal:01} have performed a survey of LMXBs carried out with 
the \emph{ASCA} satellite. The best-fit model used by these authors to fit the 
spectrum of Ser~X-1 was a blackbody plus a cutoff power-law with a Gaussian iron 
line. 
\cite{Ooster.etal:01} have analyzed two simultaneous observations 
of this source collected with BeppoSAX and RXTE.  The authors fitted  the
broad-band (0.1-200 keV) BeppoSAX spectrum with a model consisting of a
disk blackbody, a reflection component described by the XSPEC model 
{\texttt pexrav}, and a Gaussian line. 
However, in that case the improvement in $\chi^2$ with respect to a model 
consisting of a blackbody, a Comptonization spectrum modeled by compST, 
and a Gaussian was not significant, and therefore it was not possible to 
draw any definitive conclusion about the presence of a reflection continuum. 

\citet{Bhatta.etal:07} carried out the analysis of three 
\emph{XMM-Newton} observations of this system. They managed to fit the EPIC/pn 
spectrum with a model consisting of disk blackbody, a Comptonization continuum 
modeled with {\texttt compTT} and a {\texttt diskline}, i.e. a Gaussian line 
distorted and smeared by the Keplerian velocity field in the accretion disk
\citep{Fabian.etal:89}.
They found strong evidence that the Fe line 
has an asymmetric profile 
and therefore that the line originates from reflection in the inner rim of 
the accretion disk. Fitted with a Laor profile \citep{Laor.etal:91}, the 
line shape gave an inner disk radius of $4-5 \, R_g$ or $16 \, R_g$ (depending 
from the observation) and an inclination angle to the binary system of 
$40-50^\circ$. \cite{Cackett.etal:08}, from data collected by 
\emph{SUZAKU}, performed a study of the iron line profiles in a 
sample of three LMXBs including Ser~X-1. From the analysis of XIS and PIN 
spectra, they found a good fit of the broad-band continuum using a 
blackbody, a disk blackbody and a power-law.
Two years later \cite{Cackett.etal:10} re-analyzed \emph{XMM-Newton} 
and  \emph{SUZAKU} data of a sample of 10 LMXBs that includes Ser~X-1, 
focusing on the iron line - reflection emission. In particular, for 
Ser~X-1, they analyzed 4 spectra: three Epic-PN spectra 
obtained with \emph{XMM-Newton} and one obtained with the XIS and 
the PIN instruments on board of \emph{SUZAKU}. Initially, they 
fitted the spectra of the continuum emission using a 
phenomenological model, consisting of a blackbody, a disk-blackbody 
and a power-law. Then, they started the study of the Fe line adding 
first a diskline component and after a reflection component convolved 
with \texttt{rdblur} (that takes into account smearing effects due 
to the motion of the emitting plasma in a Keplerian disk). 
They obtained different results for the smearing parameters both for 
different observations and for different models used on the same observation.
For sake of clarity these results are summarized in Table \ref{tab:intro}.

\citet{Miller.etal:13} analyzed two NuSTAR observations 
carried out on July 2013. 
They fitted the continuum emission using a model consisting of 
a blackbody, a disk blackbody and a power-law. With respect to this 
continuum model, evident residuals were present around 
6.40-6.97 keV, suggesting the presence of a Fe line. Therefore they 
added a \texttt{kerrdisk} component to the continuum to fit the emission
line, taking into account a possible non-null spin parameter for the neutron
star. 
They also tried to fit the reflection spectrum (i.e.\ the iron line and 
other expected reflection features) with the self-consistent reflection 
model \texttt{reflionx}, a modified version of reflionx calculated for a 
blackbody illuminating spectrum, convolved with the \texttt{kerrconv} component.
The addition of the reflection component gave a significant improvement
of the fit. In most cases the best fit gave low inclination angles 
(less than $\sim 10^\circ$), in agreement with recent optical observations
\citep{Cornelisse.etal:13}, inner disk radii compatible with the Innermost 
Stable Circular Orbit (ISCO), corresponding to about 6 Rg for small values 
of the spin parameter, a ionization parameter $\log \xi \sim 2.3-2.6$, 
and a slight preference for an enhanced iron abundance.
The fit resulted quite insensitive to the value of the adimensional 
spin parameter, a, of the neutron star.

More recently, \citet{Chiang.etal:16} analysed a recent 300 ks Chandra/HETGS 
observation of the source performed in the "continuous clocking" mode and thus 
free of photon pile-up effects. They fitted the continuum with a combination
of multicolor disk blackbody, blackbody and power-law. The iron line was found
significantly broader than the instrumental energy resolution and fitting 
this feature with a diskline instead of a broad Gaussian gave a significant
improvement of the fit. They also tried self-consistent reflection models,
namely the reflionx model with a power-law continuum as illuminating source
and xillver \citep[see e.g.][]{Garcia.etal:13},
to describe the iron line and other reflection features, yielding consistent
results. In particular, this analysis gave a inner radius of $\sim 7-8$ R$_g$ 
and an inclination angle of about 30 deg.


As described above, different continuum models were used to fit the 
spectrum of Ser X-1 observed with various instruments at different times. 
In Table \ref{tab:intro} we summarize the results of the spectral 
analysis of this source obtained from previous studies, and in particular
the results obtained for the iron line and the reflection model. 
Quite different values have been reported for the inclination angle (from less
than 10 deg to about 40 deg), for the inner disk radius (from 4 to more 
than 100 R$_g$)
and for the iron line centroid energy and/or the ionization parameter 
$\log \xi$ indicating that the disk is formed by neutral or very highly 
ionized plasma.

In this paper we re-analyzed all the available public \emph{NuSTAR} 
observations of Ser X-1, fitting the iron line and other reflection features 
with both phenomenological and self-consistent reflection models. 
These data were already analysed by \citet{Miller.etal:13} using 
a different choice of the continuum and reflection models.
We compare these results with those obtained from three \emph{XMM-Newton} 
observations \citep[already analyzed by][]{Bhatta.etal:07} fitted with the 
same models. 
We choose to re-analyse \emph{NuSTAR} and \emph{XMM-Newton}
spectra because these instruments provide the largest effective area
available to date, coupled with a moderately good energy resolution, at 
the iron line energy, and a good broad-band coverage. Moreover, the source
showed similar fluxes during the \emph{NuSTAR} and \emph{XMM-Newton} observations.
Note also that \emph{NuSTAR} is not affected by pile-up problems in the whole
energy range. 
The spectral results obtained for \emph{NuSTAR} and \emph{XMM-Newton}
are very similar to each other and the smearing parameters of the 
reflection component are less extreme than those found by \citet{Miller.etal:13},
and in good agreement with the results obtained from the Chandra observation 
\citep{Chiang.etal:16}.
In particular we find an inner disk radius in the range $10-15\, R_g$ and
an inclination angle with respect to the line of sight of $25-30^\circ$.

\section{Observations and Data Reduction}
\label{sec:reduc} 
In this paper we analyze data collected by the \emph{NuSTAR} 
satellite. Ser~X-1 has been observed twice with \emph{NuSTAR}, obsID: 
30001013002 (12-JUL-2013) and obsID: 30001013004 (13-JUL-2013). 
The exposure time of each observation is about 40~ksec. The data 
were extracted using NuSTARDAS (NuSTAR Data Analysis Software) v1.3.0. 
Source data have been extracted from a circular region with 120" 
radius whereas the background has been extracted from a circular 
region with 90" radius in a region far from the source. First, we 
run the "nupipeline" with default values of the parameters 
as we aim to get "STAGE 2" events clean. Then spectra for both 
detectors, FPMA and FPMB, were extracted using the "nuproducts" 
command. Corresponding response files were also created as output 
of nuproducts. A comparison of the FPMA and FPMB spectra, indicated 
a good agreement between them. To check this agreement, we 
have fitted the two separate spectra with all parameters tied to each 
other but with a constant multiplication factor left free to vary. 
Since the value of this parameter is $1.00319 \pm 0.00145$, our 
assumption is basically correct.
Following the same approach described in \citet{Miller.etal:13}, 
we have therefore created a single 
added spectrum using the "addascaspec" command. A single response 
file has been thus created using "addrmf", weighting the two 
single response matrices by the corresponding exposure time. In this way,
we obtain a summed spectrum for the two \emph{NuSTAR} observations 
and the two \emph{NuSTAR} modules. 
We fitted this spectrum in the 3-40 keV energy range, where the emission 
from the source dominates over the background. 

We have also used non-simultaneous data collected with XMM-Newton satellite 
on March 2004. The considered obsID are 0084020401, 0084020501  and 
0084020601. All observations are in Timing Mode and each of them has a 
duration of $\sim$ 22 ksec.
We extracted source spectra, background spectra and response 
matrices using the SAS (Science Analysis Software) v.14 setting the 
parameters of the tools accordingly. 
We produced a calibrated photon event file using reprocessing tools 
"epproc" and "rgsproc" for PN and RGS data respectively.
We also extracted the MOS data; these were operated in uncompressed timing 
mode. However, the count rate registered by the MOS was in the range 
$290-340$ c/s, which is above the threshold for avoiding deteriorated response 
due to photon pile-up. The MOS spectra indeed show clear signs of pile-up 
and we preferred not to include them in our analysis, since these detectors 
cover the same energy range of the PN.

Before extracting the spectra, we filtered out contaminations due
to background solar flares detected in the 10-12 keV Epic PN light-curve.
In particular we have cut out about 600 sec for obsID 0084020401, 
about 800 sec for obsID 0084020501 and finally about 1600 sec for
obsID 0084020601. In order to remove the flares, we applied time 
filters by creating a GTI file with the task "tabgtigen". 
In order to check for the presence of pile-up  we have run the 
task "epatplot" and we have found significant contamination in each 
observation. The count-rate registered in the PN observations was 
in the range 860-1000 c/s that is just above the limit for avoiding 
contamination by pile-up. Therefore, we extracted the source spectra 
from a rectangular region (RAW X$\geq$26) and (RAW X$\leq$46) 
including all the pixels in the y direction but excluding the 
brightest columns at RAW X = 35 and RAW X = 36. This reduced 
significantly the pile up  (pile up fraction below a few percent 
in the considered energy range).
 
We selected only events with PATTERN $\leq$ 4 and FLAG$=$0 that 
are the standard values to remove spurious events. 
We extracted the background spectra from a similar region to the one 
used to extract the source photons but in a region away from the 
source included between (RAWX$\geq$1) and (RAWX$\leq$6 ). Finally,
for each observation, using the task 'rgscombine' we have obtained 
the added source spectrum RGS1+RGS2, the relative added background 
spectrum along with the relative response matrices. We have fitted 
RGS spectrum in the 0.35-1.8 keV energy range, whereas the Epic-PN  
in the 2.4-10 keV energy range.

Spectral analysis has been performed using XSPEC v.12.8.1 \citep{Arnaud:96}.
For each fit we have used the \texttt{phabs} model in XSPEC to 
describe the neutral photoelectric absorption due to the interstellar medium with 
photoelectric cross sections from \citet{Verner.etal:96} and element abundances 
from \citet{Wilms.etal:00}. For the \emph{NuSTAR} spectrum, which lacks of low-
energy coverage up to 3 keV, we fixed the value of the equivalent hydrogen column,
$N_H$, to the same value adopted by \citet{Miller.etal:13}, namely 
N$_{H}\, = 4 \times 10^{21}$ cm$^{-2}$ \citep{Dickey.etal:90}, while for the 
XMM-Newton spectrum we left this parameter free to vary in the fit, finding a 
slightly higher value (see Tab.\ \ref{tab:fit_1} and \ref{tab:fit_2}).
As a further check, we have fitted the NuSTAR spectrum fixing N$_H$ to the same
value found for the XMM spectrum, but the fit parameters did not change
significantly.



\section{Spectral Analysis}

\subsection{\emph{NuSTAR} spectral analysis}
\label{sec:spec} 

The \emph{NuSTAR} observations caught the source in a high-luminosity
($\sim 10^{38}$ erg/s, \citet{Miller.etal:13}) state, therefore 
most probably in a soft state. 
As seen in other similar atoll sources, the spectrum of Ser~X-1 is 
characterized by a soft component (i.e.\ blackbody), interpreted as
thermal  emission from the accretion disk, a hard component (i.e.\ 
a Comptonization spectrum), interpreted as saturated Comptonization 
from a hot corona, and often by the presence of a broad iron emission 
line at $6.4-6.97$ keV depending on the iron ionization state. We 
used the Comptonization model \texttt{nthComp} \citep{Zycki.etal:99} 
in XSPEC, with a blackbody input seed photon spectrum, 
to fit the hard component. 
We used a simple blackbody to describe the soft component. 
Substituting the blackbody with a multicolor disk blackbody, 
\texttt{diskbb} in XSPEC, gives a similar quality fit and the 
best-fit parameters do not change significantly.  

\begin{figure*}
\centering

    \includegraphics[angle=-90,width=7.6cm]{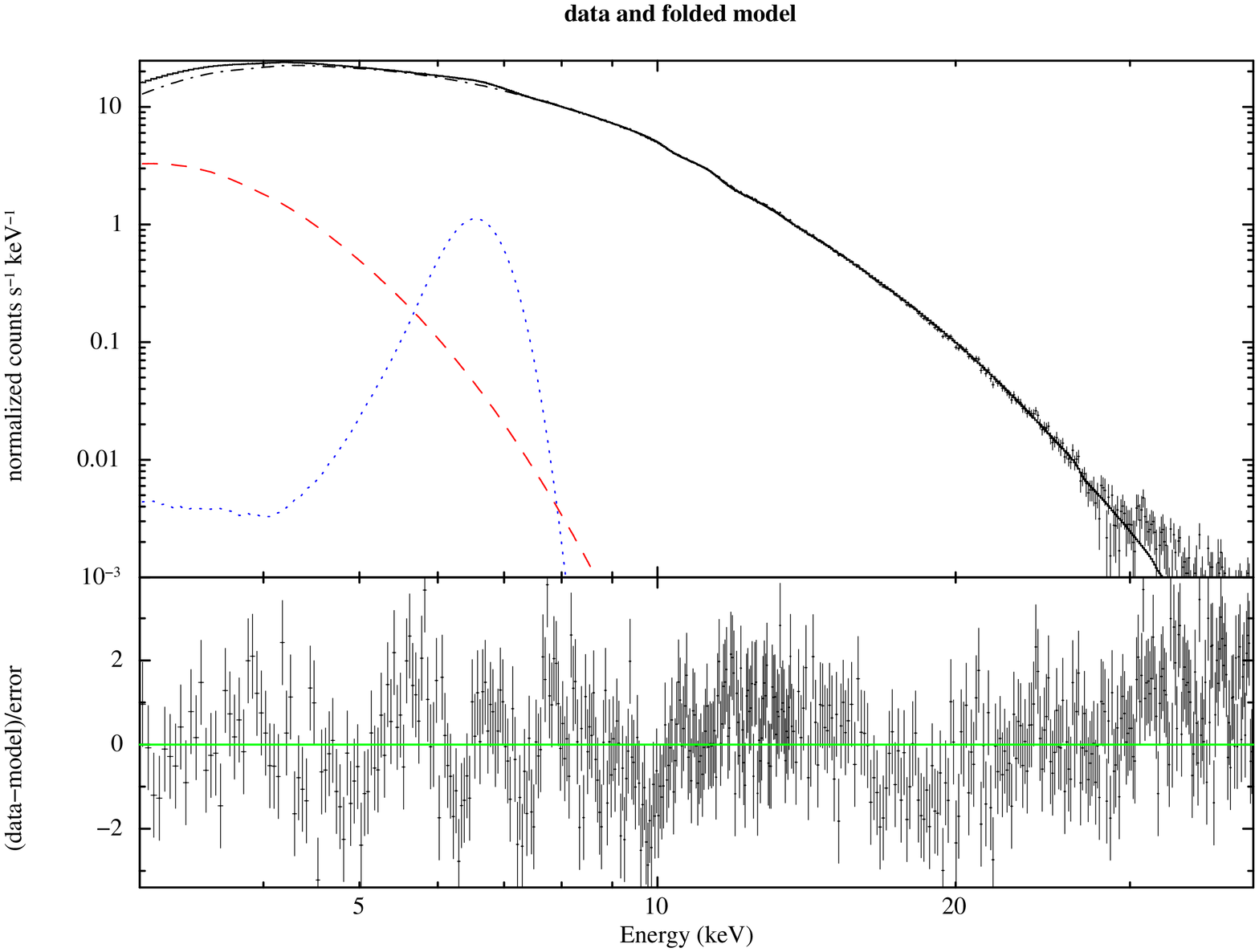}
  \includegraphics[angle=-90,width=7.6cm]{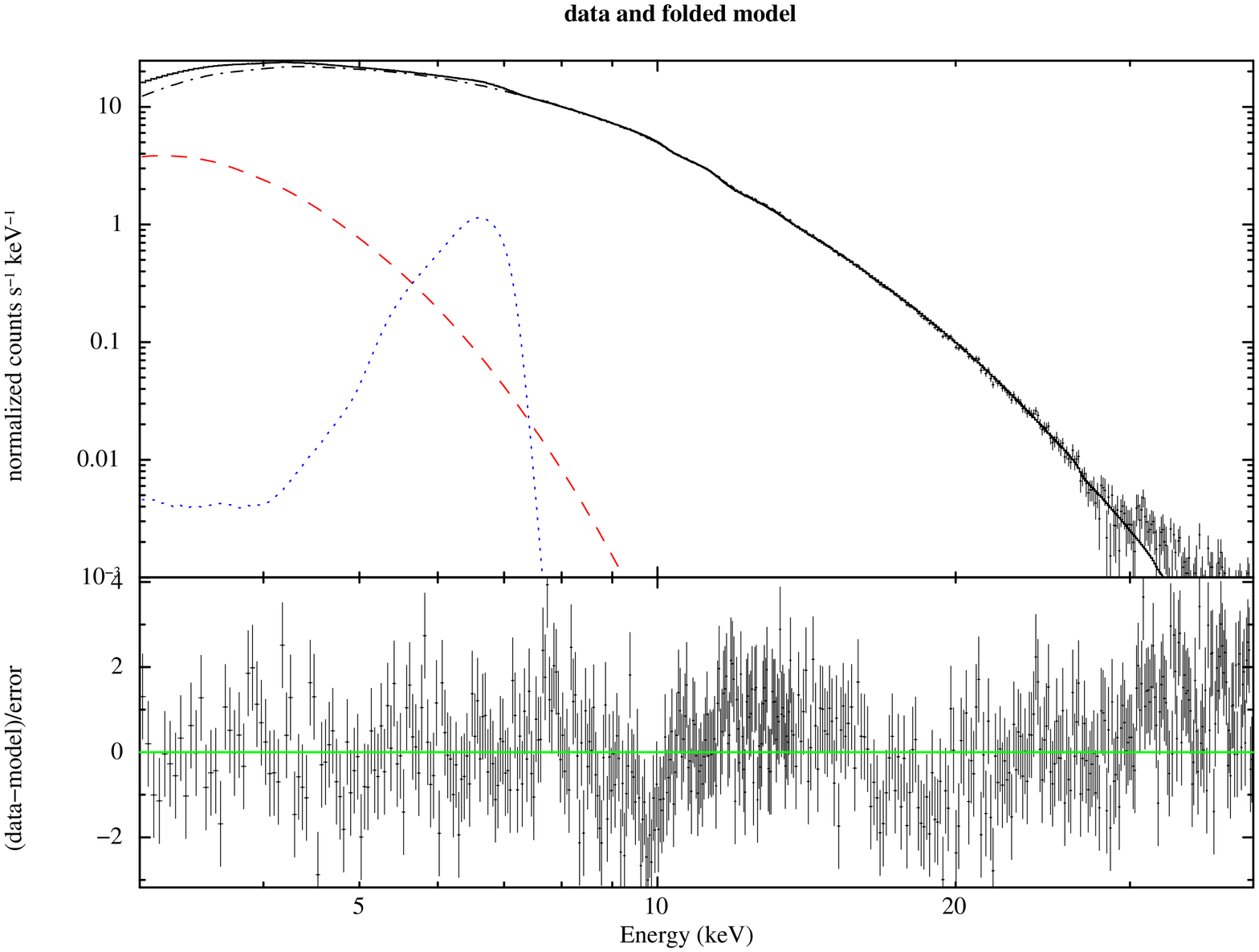}
  \includegraphics[angle=-90,width=7.6cm]{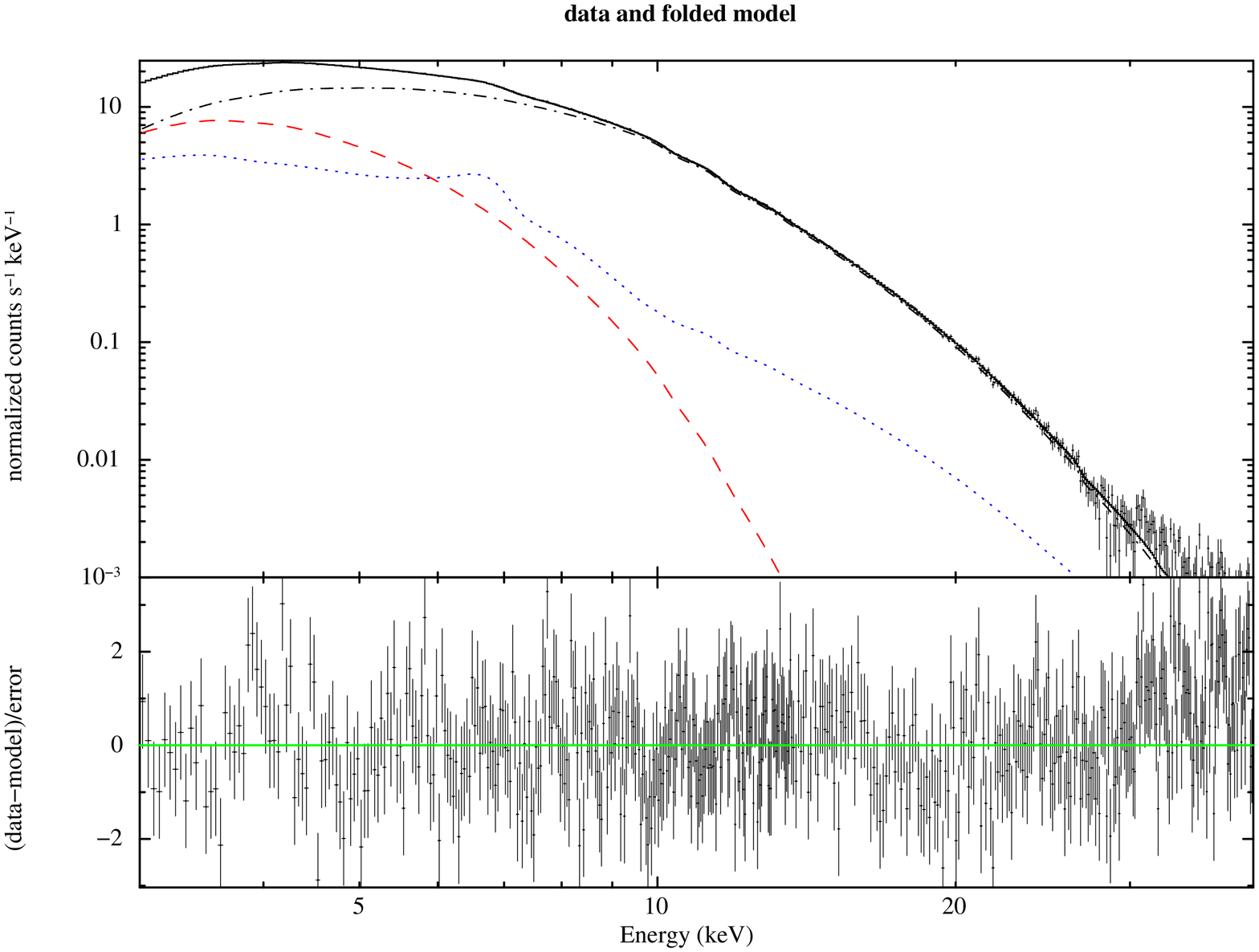}
  \includegraphics[angle=-90,width=7.6cm]{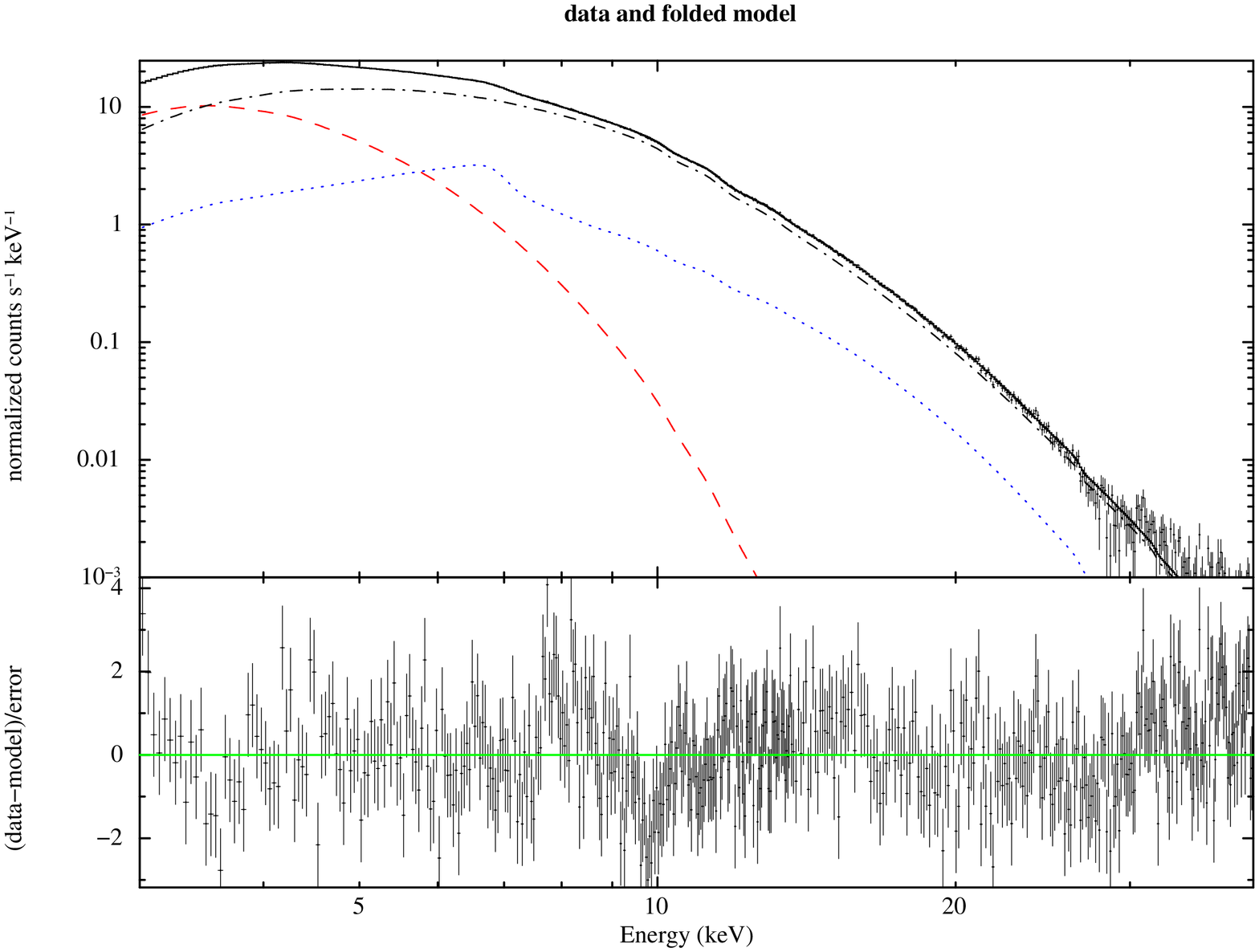}
 
\caption{NuSTAR spectra of Ser X-1 and best-fitting model together with residuals in units of sigma for the corresponding model. These are: \emph{Top left}: 'gauss' --- \emph{Top right}: 'diskline' --- \emph{Bottom left}: 'rdb-reflio' --- \emph{Bottom right}: 'rdb-rfxconv'. Dashed lines indicate the black-body component, dotted lines indicate the reflection components (i.e. the Gaussian or Diskline profile for the iron line, top panels, or the self-consistent reflection component, bottom panels, respectively), and the dashed-dotted lines indicate the comptonized component.}
     
   \label{fig:fit_1}
\end{figure*}

To fit the iron line we first tried simple models such as a Gaussian 
profile or a \texttt{diskline} \citep{Fabian.etal:89}. 
The best-fit parameters, obtained using alternatively a Gaussian 
or diskline profile, are in good agreement with each other (see Tab.\
\ref{tab:fit_1}). 
Using a \texttt{diskline} instead of a Gaussian profile we get an improvement 
of the fit corresponding to $\Delta \chi^2 = 54$ for the addition of two parameters. 
Spectra, along with the best-fit model and residuals are shown in 
Fig.\ref{fig:fit_1}. In both cases, the fit results are poor (the relative 
null hypothesis probability is $2.8 \times 10^{-8}$; the reduced $\chi^{2}\,$ 
are still relatively large, and evident residuals are present, especially 
above 10 keV, see Fig.\ref{fig:fit_1}).

In order to fit the residuals at high energy, we added a \texttt{powerlaw} 
component (a hard tail) to all the models described above. A hard  
power-law tail is often required to fit high-energy residuals of 
atoll sources in the soft state \citep[see e.g.][]{Pintore.etal:15, 
Pintore.etal:16, Iaria.etal:01, Iaria.etal:02}, and this component may 
also be present in the spectrum of Ser X-1 \citep[see][]{Miller.etal:13}. 
Unless it is specified otherwise, 
for every fit, we froze the power-law photon index to the value 
found by \citet{Miller.etal:13} for Ser X-1, that is 3.2. 
The new models are now called \emph{gauss-pl} and \emph{diskline-pl}, 
respectively. The new best fit parameters are reported in Tab \ref{tab:fit_1}. 
While the best-fit parameters do not change significantly with the addition
of this component, we get an improvement of the fit corresponding to a
reduction of the $\chi^2$ by $\Delta \chi^2 = 123$ (for the model with a 
Gaussian line profile) and $\Delta \chi^2 = 113$ (for the model with a 
diskline profile) for the addition of one parameter, respectively. 
The probabilities of chance improvement of the fit are $8.5 \times 10^{-24}$ 
and $8.6 \times 10^{-23}$, respectively.  Some residuals are 
still present between 10 and 20 keV probably caused by the presence of an 
unmodeled Compton hump.
Note that the soft blackbody component remains significant even after the
addition of the power-law component. If we eliminate this component 
from the fitting model we get a worse fit, corresponding to a decrease by 
$\Delta \chi^2 = 245$ for the addition of two parameters when the soft 
component is included in the fit and a probability of chance improvement
of the fit of $\sim 3 \times 10^{-44}$.

\subsection{Reflection models}
\label{sec:refl} 
We have also tried to fit the \emph{NuSTAR} spectrum of Ser X-1 with 
more sophisticated reflection models, performing a grid of fit with 
self-consistent models such as \texttt{reflionx} or \texttt{rfxconv}.
Reflionx and rfxconv models both include the reflection continuum, the
so called Compton hump caused by direct Compton scattering of the 
reflected spectrum, and discrete features (emission lines and absorption 
edges) for many species of atoms at different ionization stages 
\citep{Ross.etal:05, Kolehmainen.etal:11}.  

The \texttt{reflionx} model depends on 5 parameters, that are 
the abundance of iron relative to the solar value, the photon index of 
the illuminating power-law spectrum ($\Gamma$, ranging between 1.0 to 3.0), 
the normalization of reflected spectrum, the redshift of the source, 
and the ionization parameter $\xi = L_X / (n_e r^2)$ where $L_X$ is the 
X-ray luminosity of the illuminating source, $n_e$ is the electron density 
in the illuminated region and $r$ is the distance of the illuminating 
source to the reflecting medium. 
When using \texttt{reflionx}, which uses a power-law as illuminating 
spectrum, in order to take into account the high-energy roll over of the 
Comptonization spectrum, we have multiplied it by a 
high-energy cutoff, \texttt{highecut}, with the folding energy 
E$_{fold}$ set to 2.7 times the electrons temperature kT$_{e}$ 
and the cutoff energy E$_{cutoff}$ tied to 0.1 keV. In this 
way we introduce a cutoff in the reflection continuum, which otherwise 
resembles a power-law. The cut-off energy fixed at 2.7 times the 
electron temperature of the Comptonization spectrum (assumed to be 
similar to a blackbody spectrum), is appropriate for a saturated 
Comptonization \citep[see e.g.][]{Egron.etal:13}. To fit the 
Comptonization continuum we used the \texttt{nthComp} model.  
Moreover we fixed the photon index of the illuminating spectrum, 
$\Gamma$, to that of the \texttt{nthComp} component.  
We stress  that in our analysis we use a different 
\texttt{reflionx} reflection model with respect to that used by 
\citet{Miller.etal:13}. In fact we used a model  that assumes an 
input power-law spectrum as the source of the irradiating flux modified, 
in order to mimic the nthcomp continuum, by introducing the model 
component \texttt{highecut}. \citet{Miller.etal:13} instead used a 
modified version of reflionx calculated for a blackbody input spectrum, 
since that component dominates their phenomenological continuum.

\texttt{rfxconv} is an updated version of the code in \cite{Done.etal:06}, 
using \cite{Ross.etal:05} reflection tables. 
This is a convolution model that can be used with any input continuum 
and has therefore the advantage to take as illuminating spectrum the
given Comptonization continuum.  It depends on 5 parameters: 
the relative reflection fraction (rel-refl defined as $\Omega/2\pi$,
namely as the solid angle subtended by the reflecting disk as seen from
the illuminating corona in units of $2\pi$), the cosine of the inclination 
angle, the iron abundance relative to the Solar value, the ionization 
parameter Log $\xi$ of the accretion disk surface, and the redshift of 
the source. 

Due to its high velocities, the radiation re-emitted from the plasma 
located in the inner accretion disk undergoes to Doppler and relativistic 
effects (which smears the whole reflection spectrum). In order to take 
these effects into account we have convolved the reflection models
with the \texttt{rdblur} component (the kernel of the diskline model),
which depends on the values of the inner and outer disk radii, in units
of the Gravitational radius ($R_g = G M/c^2$), the inclination angle of the 
disk (that was kept tied to the same value used for the reflection model),
and the emissivity index, Betor, that is the index of the power-law dependence 
of the emissivity of the illuminated disk (which scales as $r^{Betor}$). 
Finally, we have also considered the possibility that neutron star has a 
spin. In this case, the reflection component has been convolved
with the \texttt{Kerrconv} component \citep{Brenneman.etal:06} that through 
its adimensional spin parameter 'a' allowed us to implement a grid of models 
exploring different values of 'a' (see Appendix \ref{sec:appen}). 
For this model there is also the possibility to fit the emissivity 
index of the inner and outer part of the disk independently, although in our 
fits we used the same emissivity index for the whole disk.
For all the fits we have fixed the values of R$_{out}$ to 2400 $R_g$,
the iron abundance to solar value, Fe/solar = 1, and the redshift of the
source to 0. 
The best fit parameters are reported in Tab \ref{tab:fit_1}--\ref{tab:fit_4}.

We started to fit the data adding a reflection component, \texttt{reflionx} 
or \texttt{rfxconv}, convolved with the blurring component \texttt{rdblur}, 
to the continuum model given by the blackbody and the nthcomp components
(models are called \emph{rdb-reflio} and \emph{rdb-rfxconv}, respectively). 
Fit results for both models are acceptable, with $\chi_{red}^{2}\,$ close to 1.09. 
There are a few differences between the best-fit parameters of the rdb-reflio 
model with respect to those of the rdb-rfxconv model. In particular the 
\emph{rdb-rfxconv} model gives a lower value of R$_{in}$, while the 
\emph{rdb-reflio} model gives a higher ionization parameter (although 
with a large uncertainty).
Spectra, along with the best-fit model and residuals are reported in 
Fig.\ref{fig:fit_1}. The residuals that are very similar
for the two models, apart for the 8-10 keV energy range where 
\emph{rdb-reflio} model shows flatter residuals than 
\emph{rdb-rfxconv} model (see Fig.\ \ref{fig:fit_1}).

\begin{figure*}
\centering

   \includegraphics[angle=-90,width=7.6cm]{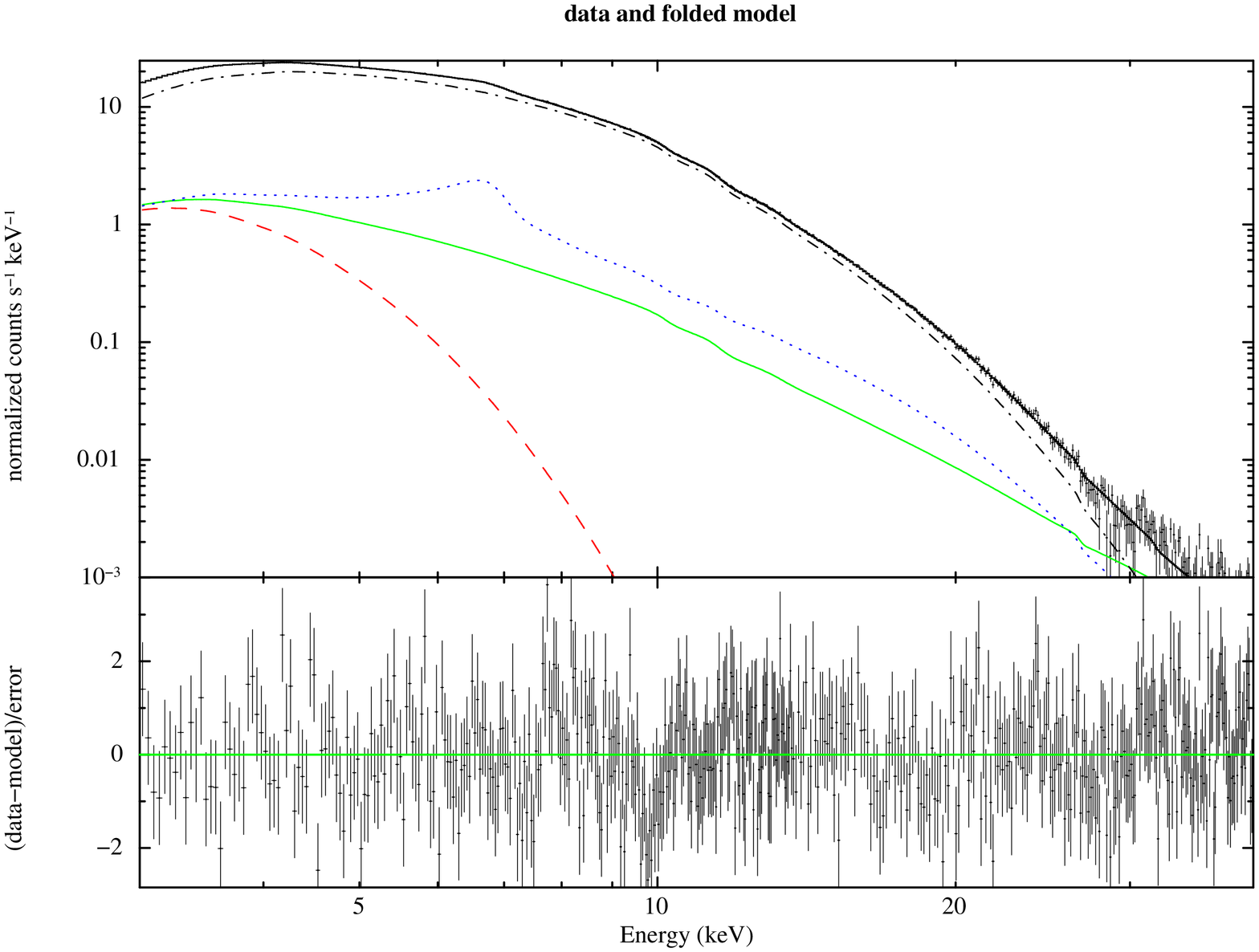}
  \includegraphics[angle=-90,width=7.6cm]{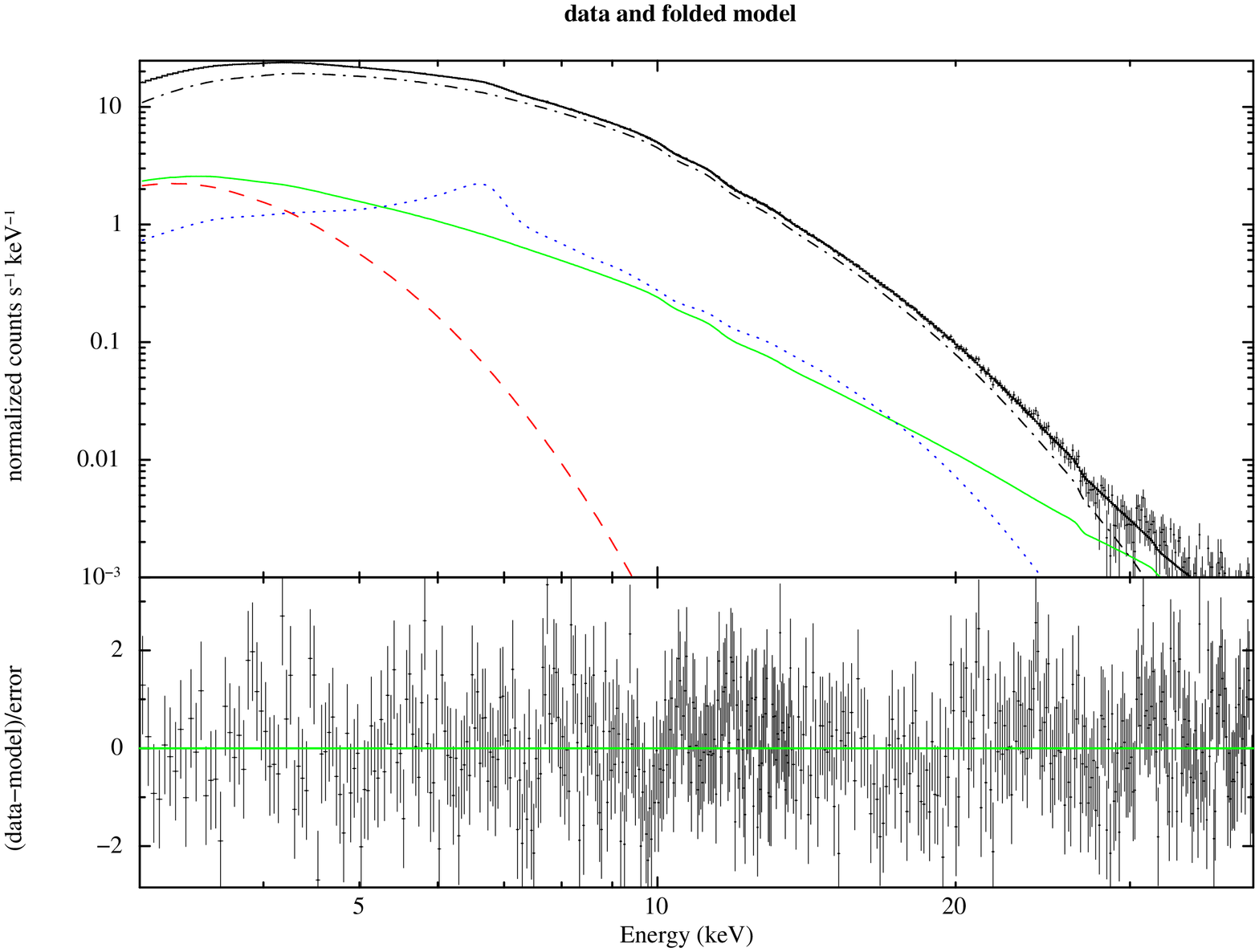}
   \includegraphics[angle=-90,width=7.6cm]{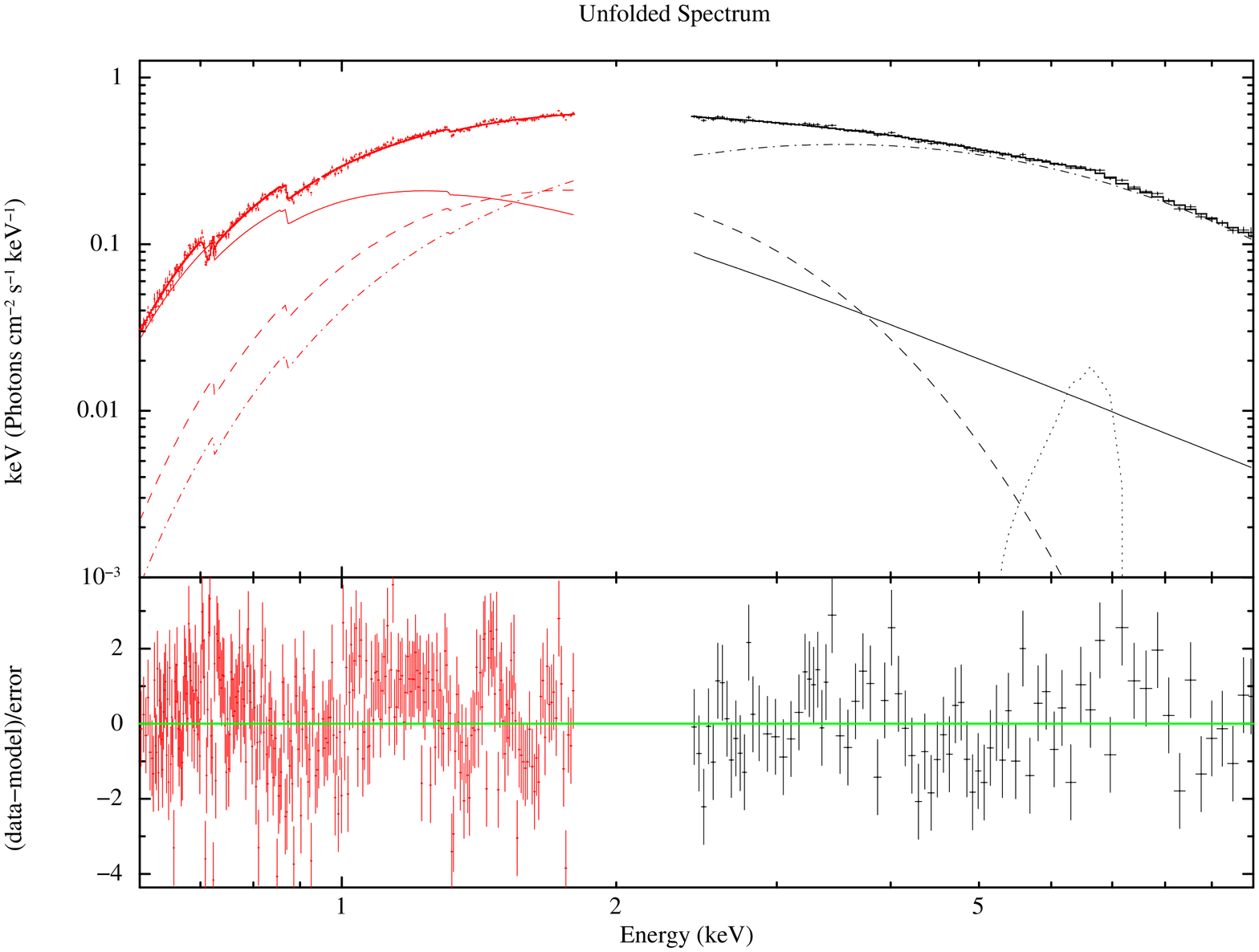}
  \includegraphics[angle=-90,width=7.6cm]{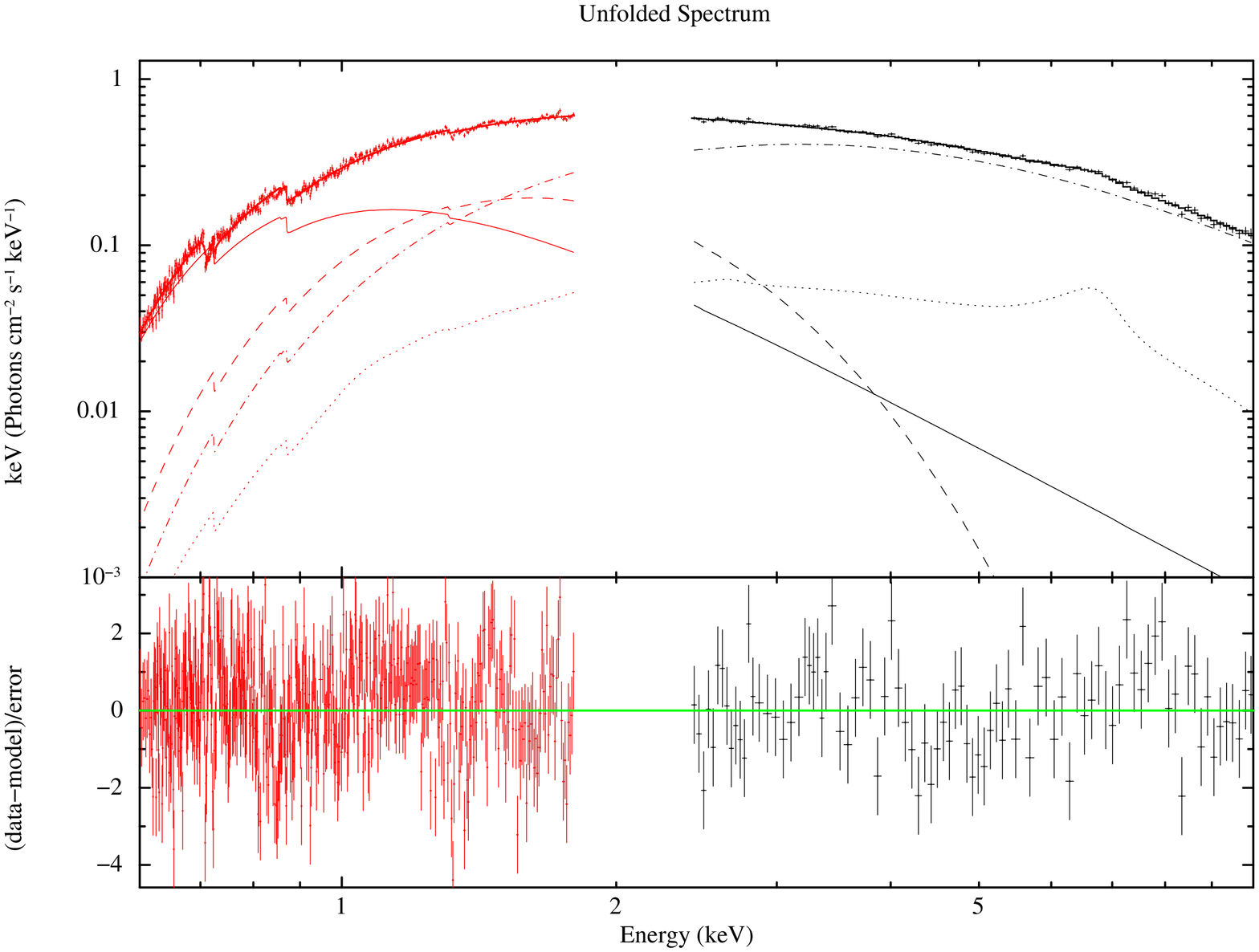}

\caption{\emph{Top panels}: NuSTAR spectra of Ser X-1 and best-fitting model together with residuals in units of sigma for the corresponding model. These are: \emph{Top left}: 'rdb-reflio-pl' --- \emph{Top right}: 'rdb-rfxconv-pl'. {Bottom panels}: XMM-Newton spectra and best-fitting model together with residuals in units of sigma for the corresponding model. These are: \emph{Bottom left}: 'diskline-pl-xmm' --- \emph{Bottom right}: 'rdb-rfxconv-pl-xmm'. For clarity only the first XMM-Newton observation is shown. Dashed lines indicate the black-body component, dotted lines indicate the reflection components (i.e. the Diskline profile for the iron line or the self-consistent reflection component), the solid line indicates the power-law component, and the dashed-dotted lines indicate the comptonized component.}

   \label{fig:fit_2}
\end{figure*}

As  before, we also tried to add a power-law component to the models obtained 
by the convolution of the blurring component (rdblur) with the two different 
reflection components (rfxconv or reflionx). The two new models are called 
\emph{rdb-rfxconv-pl} and \emph{rdb-reflion-pl}, respectively. In both cases
we get a significant improvement of the fit, with $\Delta \chi^2 = 90$ for 
the addition of two parameters and $\Delta \chi^2 = 66$ for the addition of 
one parameter, respectively. In these cases, an F-test yields a probability of 
chance improvement of $3.1 \times 10^{-15}$ for \emph{rdb-reflion-pl} and 
$6.1 \times 10^{-19}$ for \emph{rdb-rfxconv-pl} model, respectively. 
 Spectra, along with best-fit model and residuals are reported in Fig.\ \ref{fig:fit_2}, 
whereas values of the best-fit parameters are listed in Tab.\ \ref{tab:fit_2}.
Residuals are now flat (see plots reported in upper panels of Fig.\ \ref{fig:fit_2}). 
Note also that in this way  we get more reasonable values of the best-fit parameters,
especially for the ionization parameter, $\log \xi$, which is around 2.7
for both models, in agreement with the centroid energy of the iron line at about
6.5 keV, and well below 3.7 (a ionization parameter $\log \xi \sim 3.7$ 
would imply that the matter of the accretion disk would be fully ionized). 


In summary, the best fit of the \emph{NuSTAR} spectrum of Ser X-1 is obtained
fitting the continuum with a soft blackbody component, a Comptonization 
spectrum, and a hard power-law tail and fitting the reflection features 
with the \texttt{rfxconv} model smeared by the \texttt{rdblur} component,
since the fitting results are quite insensitive to the value of the 
spin parameter a (see Appendix \ref{sec:appen}). This fit, corresponding 
to a $\chi^2(dof) = 912.5(911)$, gives a blackbody temperature of 
$\simeq 0.54$ keV, a temperature of the seed photons for the Comptonization 
of $\simeq 0.93$ keV, an electron temperature of the Comptonizing corona of 
$\simeq 2.70$ keV and a photon index of the primary Comptonized component 
of $\simeq 2.17$, whereas the photon index of the hard power-law tail is 
steeper, around 3.2.
The reflection component gives a reflection amplitude (that is the 
solid angle subtended by the accretion disk as seen from the Comptonizing 
corona) of $\simeq 0.24$ and a ionization parameter of $\log \xi \simeq 2.7$. 
The smearing of the reflection component gives an
inner disk radius of $R_{in}$ ranging between 10 and 16 $R_g$, and 
inclination angle of the disk with respect to the line of sight of 
$i \simeq 27^\circ$, and the emissivity of the disk scaling as 
$\propto r^{-2.6\pm0.2}$.
Note that the Compton hump is highly significant. To evaluate its 
statistical significance we can compare the best fit obtained with the model 
diskline-pl with the best fit given by the model rdb-rfxconv-pl (the main 
difference between the two models is in fact that rfxconv contains the 
reflection continuum and diskline does not). Using rfxconv instead of 
diskline we get a decreases of the $\chi^2$ by $\Delta \chi^2 = 87$ 
for the addition of 1 parameter and an F-test probability of chance 
improvement of $8 \times 10^{-20}$, which is statistically significant.

\subsection{XMM-Newton Spectral Analysis}
\label{sec:xmm} 

We have also carried out the analysis of \emph{XMM-Newton} observations of 
Ser~X-1. A previous study, based only on the PN data analysis, has been 
reported by \cite{Bhatta.etal:07}. We updated the analysis by performing
the fit of the RGS spectra in the 0.35--1.8 keV energy range and the PN 
spectra in the 2.4--10 keV energy range. Following the same approach used
for the analysis on \emph{NuSTAR} data, we assumed a continuum model composed 
of a blackbody, a hard power-law and the nthComp component.
In addition to the continuum components described above, we have also detected 
several discrete features present in all RGS spectra, both in absorption and in 
emission that were supposed to be of instrumental origin by \cite{Bhatta.etal:07}. 
The energies of the most intense features detected in our spectra lie between 
0.5 keV and 0.75 keV. To fit these features we have therefore added three 
additional gaussians to our model: two absorption lines at 0.528 keV 
and at 0.714 keV, respectively, and one in emission at 0.541 keV.
 The identification of these lines is not straightforward. The 0.528 keV
energy is close to the neutral O K$\alpha$ line, expected at a rest frame energy 
of 0.524 keV, while the 0.541 keV emission line is close to the expected energy 
of the O~I edge at 0.538 keV. These two lines may be therefore instrumental features 
caused by a miscalibration of the neutral O edge in the RGS. The other absorption line 
at 0.714 keV is close to the O~VII absorption edge expected at a rest-frame energy of 
0.739 keV. Given that the identification of these lines is uncertain, we will not
discuss them further in the paper. 
To this continuum we first added a diskline (model called \emph{diskline-pl-xmm},
see Table \ref{tab:fit_1}) to fit the iron line profile. Then we fitted the 
spectra substituting the diskline with the self-consistent reflection model that 
gave the best fit to the \emph{NuSTAR} data, that is 'rfxconv', convolved with 
the smearing component 'rdblur' (model called \emph{rdb-rfxconv-pl-xmm}, results 
are reported in Table \ref{tab:fit_2}). 

We have performed the fit of the spectrum obtained from these three observations 
simultaneously, tying parameters of the RGS with the all parameters of the PN from 
the same observation. The spectra of the three XMM observations are very similar 
with each other, except for the soft black body temperature that was left free to 
vary in different datasets. 
Values of the best-fit parameters of the model \emph{diskline-pl-xmm} result to be 
in good agreement with what we have found from the fit of the \emph{NuSTAR} spectra 
with the same model. 


We have also performed the fit with a model including the reflection component
\texttt{rfxconv}, called \emph{rdb-rfxconv-pl-xmm}.
As before, in order to take into account structures visible in the RGS spectra, we 
have added three gaussians to the model. As before we have tied parameters of the 
RGS to the corresponding parameters of the PN from the same observation except
for the parameter kT$_{bb}$ that was left free to vary among the three observations. 
Note also that for the these fits the inclination angle is fixed to the corresponding
values we found from the \emph{NuSTAR} spectra.
Results are reported in Table \ref{tab:fit_2}, and are in good agreement with those
obtained for the \emph{NuSTAR} spectrum.


\section{Discussion }
\label{sec:disc}

Ser X-1 is a well studied LMXB showing a broad emission line at $6.4 - 6.97$
keV interpreted as emission from iron at different ionization states and
smeared by Doppler and relativistic effects caused by the fast motion of
matter in the inner accretion disk. Moderately high energy resolution spectra 
of this source have been obtained from \emph{XMM-Newton}, \emph{Suzaku}, 
\emph{NuSTAR}, and \emph{Chandra}. 
However, as described in Sec.\ \ref{sec:intro}, spectral
results for the reflection component are quite different for different 
observations or for different models used to fit the continuum and/or the
reflection component. While spectral differences in different observations
may be in principle justified by intrinsic spectral variations of the source,
differences caused by different continuum or reflection models should be
investigated in detail in order to give a reliable estimate of the parameters
of the system. For instance, in a recent \emph{NuSTAR} observation
analyzed by \citet{Miller.etal:13}, assuming a modified version of 
\texttt{reflionx} calculated for a black-body input spectrum, the authors 
report a significant detection of a smeared reflection component in this source, 
from which they derive an inner radius of the disk broadly compatible with the 
disk extending to the ISCO (corresponding to 6 Rg in the case $a = 0$) and 
an inclination angle with respect to the line of sight $< 10^\circ$. 
On the other hand, \cite{Chiang.etal:16},   
analysing a recent 300 ks \emph{Chandra/HETGS} observation of the source 
obtained a high-resolution X-ray spectrum which gave a inner radius of 
$R_{in} \sim 7-8 \, R_g$ and an inclination angle of $\sim 30^\circ$.

In this paper we analyzed all the available \emph{NuSTAR} and 
\emph{XMM-Newton} observations of Ser X-1. These observations have been already 
analyzed by \citet{Miller.etal:13} and \citet{Bhatta.etal:07}, respectively,
who used different continuum and reflection models and report different 
results for the reflection component. The same \emph{XMM-Newton} observations
have also been analyzed by \cite{Cackett.etal:10} who also report different
results for the reflection component, with higher inner disk radii (between
15 and more than $45\, R_g$) and quite low inclinations angles ($<10^\circ$)
when using a blurred reflection model, and inclination angle between 10 and
$35^\circ$ when using a diskline component to fit the iron line profile
(see Tab.\ \ref{tab:intro} for more details).
We have shown that we can fit the \emph{NuSTAR} and \emph{XMM-Newton} spectra
independently with the same continuum model and with a phenomenological model
(i.e.\ diskline) or a self-consistent reflection model (i.e.\ reflionx or 
rfxconv) for the reflection component, finding in all our fit similar 
(compatible within the associated uncertainties) smearing parameters for 
the reflection component.

To fit these spectra we have used a continuum model composed by a blackbody 
component (bbody) and a comptonization continuum (nthcomp), which has been
widely used in literature to fit the spectra of neutron star LMXBs both in
the soft and in the hard state \citep[see e.g.][]{Egron.etal:13}. 
With respect to the continuum model used by \citet{Miller.etal:13} we have
substituted one of the two blackbody components, the hottest one, 
with a Comptonization spectrum. Since this component gives the most important 
contribution to the source flux, especially above 5 keV, we have 
subsequently used this component as the source of the reflection spectrum.
In all our fit the addition of a hard power-law component, with a photon
index $\sim 3$ significantly improved the fit. The presence of a hard 
power-law component is often found in the spectra of bright LMXBs in
the soft state \citep[see e.g.][]{Piraino.etal:07, Pintore.etal:15, 
Pintore.etal:16}, and has been interpreted as comptonization
of soft photons off a non-thermal population of electrons 
\citep[see e.g.][]{DiSalvo.et:00}. 

To fit the reflection component, which is dominated by a prominent iron
line, we have first used a phenomenological model consisting of a Gaussian
line or a diskline, with a diskline providing a better fit than a Gaussian
profile (cf.\ fitting results reported in Table \ref{tab:fit_1}).
All the diskline parameters obtained from the fitting of the \emph{NuSTAR} 
and \emph{XMM-Newton} spectra are compatible with each other, except for the 
line flux which appears to be lower during the \emph{XMM-Newton}
observations.

In order to fit the reflection spectrum with self-consistent models,
which take into account not only the iron line but also other 
reflection features, we have used both \texttt{reflionx} and 
\texttt{rfxconv} reflection models. In both these models, emission
and absorption discrete features from the most abundant elements 
are included, as well as the reflected continuum. 
We have convolved the reflection spectrum with the relativistic smearing 
model \texttt{rdblur}, taking into account Doppler and relativistic effects 
caused by the fast motion of the reflecting material in the inner accretion 
disk. We have also investigated the possibility that the neutron star has
a significant spin parameter. We have therefore performed a grid of 
fits using the \texttt{kerrconv} smearing model, instead of rdblur, 
freezing the spin parameter 'a' at different values: 0, 0.12, 0.14 and 
letting it free to vary in an additional case (see Appendix \ref{sec:appen} 
for more details).  
In agreement with the results reported by \citet{Miller.etal:13} 
we find that the fit is almost insensitive to the spin parameter but
prefers low values of the spin parameter ($a < 0.04$).

The results obtained using \texttt{reflionx} or \texttt{rfxconv} are
somewhat different in the fits not including the hard power-law component.
However, the reflection and smearing parameters become very similar 
when we add this component to the continuum model (cf.\ results in Tabs.\
\ref{tab:fit_2}, \ref{tab:fit_3}, \ref{tab:fit_4}). The addition of 
this component also significantly improves all the fits. 
We consider as our best fit model the one including the hard power-law
component, \texttt{rfxconv} as reflection component smeared by the 
\texttt{rbdblur} component (model named \texttt{rdb-rfxconv-pl} in Tab.\ 
\ref{tab:fit_2}). The fit of the \emph{XMM-Newton} spectra with the same
model gave values of the parameters that overall agree with those obtained
fitting the \emph{NuSTAR} spectra. In this case, we have found values of 
the ionization parameter log($\xi$) ranging between 2.58 and 2.71 (a bit 
higher, around 3, for the \emph{XMM-Newton} spectra) and reflection 
amplitudes between 0.2 and 0.3, indicating a relatively low superposition 
between the source of the primary Comptonization continuum and the disk 
(a value of 0.3 would be compatible with a spherical geometry of a compact 
corona inside an outer accretion disk). For the smearing parameters of the
reflection component we find values of the emissivity index of the disk 
ranging from -2.8 to -2.48, an inner radius of the disk from 10.6 to 
$16.2 \, R_g$, and an inclination angle of the system with respect to the 
line of sight of $25-30^\circ$. In our results the inclination angle is 
higher than that found by \citet{Miller.etal:13} (who report an inclination 
angle less than $10^\circ$), but is very similar to that estimated from
Chandra spectra ($25-35^\circ$, see  \cite{Chiang.etal:16}. 
Moreover, the inner disk radius we find is not compatible with the ISCO. 
Assuming a $1.4 \, M_\odot$ for the neutron star, the inner radius of the 
disk is located at $22-34$ km from the neutron star center. Note that this 
value is compatible to the estimated radius of the emission region of the 
soft blackbody component, which is in the range $19-31$ km. We 
interpret this component as the intrinsic emission from the inner disk since
this is the coldest part of the system and because 
the temperature of the blackbody component appears to be too low to represent 
a boundary layer. \\


\section{Conclusions}


The main aim of this paper is to test the robustness of disk parameters inferred from 
the reflection component in the case of neutron star LMXBs; to this aim we used 
broad-band, moderately high resolution spectra of Serpens~X-1, a neutron star LMXB 
of the atoll type with a very clear reflection spectrum that has been studied with 
several instruments. In particular, we have carried out a broad-band spectral 
analysis of this source using data collected by \emph{NuSTAR} and \emph{XMM-Newton} 
satellites, which have the best sensitivity at the iron-line energy. 
These data have been already analyzed in literature. In particular 
\citet{Miller.etal:13} have analyzed the \emph{NuSTAR} spectra and have obtained 
a low inclination angle of about 8$^\circ$, an inner disk radius compatible 
with the ISCO, a ionization parameter $\log \xi$ between 2.3 and 2.6 along with an 
iron abundance of about 3. 

In the following we summarize the results presented in this paper:

\begin{itemize}

\item We have performed the fitting using 
slightly different continuum and reflection models with respect to that used 
by other authors to fit the X-ray spectrum of this source. Our the best fit 
of the \emph{NuSTAR} spectrum of Ser X-1 is obtained fitting the continuum with 
a soft blackbody, a Comptonization spectrum, a hard power-law tail in addition 
to the reflection features. To fit the reflection features present in
the spectrum we used both empirical models and self-consistent reflection 
components as \texttt{reflionx} and \texttt{rfxconv}, as well as two 
different blurring components that are \texttt{rdblur} and\texttt{kerrconv}. 
From the analysis carried out using kerrcov we have obtained that  our fit is 
insensitive to the value assumed by the adimensional spin parameter 'a', 
in agreement with what is found by \citet{Miller.etal:13} in their analysis.

\item With regard the reflection features, we obtain consistent results
using phenomenological models (such as diskline) or self-consistent models
to fit the \emph{NuSTAR} spectrum of the source. 
In particular, the reflection component gives a reflection amplitude of 
$\Omega / 2 \pi \sim 0.2-0.3$ (where $\Omega$ is the solid angle of the disk
as seen from the corona in units of $2 \pi$) and a ionization parameter of 
$\log (\xi) \sim 2.6-2.7$. 
The smearing of the reflection component gives an inner disk radius of 
$R_{in} \sim 10.6 - 16.2\, R_{g}$, an emissivity index of the disk in the range 
$-(2.5-2.8)$, whereas the inclination angle of the disk with respect to the 
line of sight results in the range $25 - 29^\circ$. We note that the 
inner disk radius derived from the reflection component results compatible 
with the radius inferred from the soft blackbody component, which results in 
the range $19 - 31$ km. 

\item The analysis of \emph{XMM-Newton} spectra, carried out using the same 
models adopted to fit the \emph{NuSTAR} spectra, gave values of the parameters
compatible to those described above, although the two observations are
not simultaneous. The only differences are the reflection amplitude, 
$\Omega / 2 \pi \sim 0.18 - 0.19$, which results slightly lower, although still 
marginally consistent within the errors, and the ionization parameter, 
$\log (\xi) \sim 2.9 - 3.1$, which results somewhat higher with respect to the 
non-simultaneous \emph{NuSTAR} observations.

\end{itemize}

In conclusion, in this paper we performed an investigation of to which extent the 
disk parameters inferred from reflection fitting depend on the chosen spectral models
for both the continuum and the reflection component. Despite the fact that several 
authors in previous work have used basically the same continuum model, the resulting  
reflection parameters, such as the inner disk radius, $R_{in}$, and the inclination angle
are scattered over a large range of values. In this paper we have re-analyzed all the 
available public \emph{NuSTAR} and \emph{XMM-Newton} observations of Ser X-1, fitting 
the continuum with a slightly different, physically motivated model and the iron line 
with different reflection models. By performing a detailed spectral 
analysis of \emph{NuSTAR} and \emph{XMM-Newton} data of the LMXB Ser~X-1 using both 
phenomenological and self-consistent reflection models, and using a continuum model 
somewhat different from that used in literature for this source, the best fit parameters 
derived from the two spectra are in good agreement between each other. These are also
broad agreement with the findings of \citet{Miller.etal:13} although we find values of 
the inner disk and the inclination angle that are less extreme. Hence, the use of 
broad-band spectra and of self-consistent reflection models, together with an investigation
of the continuum model, are highly desirable to infer reliable parameters from the 
reflection component.

\begin{acknowledgements}
We thank the anonymous referee for useful suggestions which helped to improve the
manuscript.
The High-Energy Astrophysics Group of Palermo acknowledges support from 
the Fondo Finalizzato alla Ricerca (FFR) 2012/13, project N. 2012-ATE-0390, 
founded by the University of Palermo. This work was partially supported by 
the Regione Autonoma della Sardegna through POR-FSE Sardegna 2007-2013, 
L.R. 7/2007, Progetti di Ricerca di Base e Orientata, Project N. CRP-60529. 
We also acknowledge financial contribution from the agreement ASI-INAF I/037/12/0. 
\end{acknowledgements}


\begin{sidewaystable*}

\caption{Results of Spectral Analysis of Ser X-1 from Previous Studies \label{tab:intro}}

\scriptsize


\begin{tabular}{llcccccccccc}     
\hline\hline       

Instrument &
Continuum Model &
Reflection Model &
Line Model &
Line Energy (keV)&
Equivalent width &
R$_{in}$ (R$_{g}$) &
Incl (deg) &
Emissivity index log ($\xi$)&
Flux (ergs/cm$^{2}$/sec) &
Reference \\

\hline
\emph{ASCA}  & 
bbody+cutpowerlaw & 
--- &
gaussian &
$6.6 \pm 0.17$ &
81 eV &
--- &
--- &
--- &
--- &
Ref(1)  \\

\hline

\emph{RXTE}  & 
bbody & 
pexrav &
gaussian &
-- &
-- &
--- &
--- &
--- &
--- &
Ref(2) \\

\emph{BeppoSAX}  & 
bbody+compTT & 
--- &
gaussian &
6.46 $^{+0.12}_{-0.14}$  &
275 $^{75}_{-55}$ eV &
--- &
--- &
--- &
--- &
Ref(2) \\

\hline

\emph{XMM-Newton }  & 
diskbb+compTT & 
--- &
laor &
6.40 $^{+0.08}_{-0.00}$  &
86-105 eV &
4-16  &
40-50  &
--- &
2-10 keV: (3.3-4.2)$\times$10${^{-9}}$ &
Ref(3) \\

\hline

\emph{SUZAKU}  & 
bbody+diskbb+powerlaw & 
--- &
diskline &
6.83 $^{+0.15}_{-0.06}$  &
132$\pm$12 eV &
7.7$\pm$0.5 &
26$\pm$2 &
--- &
0.5-10 keV: 5.9$\pm$0.9)$\times$10${^{-9}}$\footnote{Estimated only for the continuum component} &
Ref(4) \\

\hline

\emph{SUZAKU}  & 
bbody+diskbb+powerlaw & 
--- &
diskline &
6.97 $^{+0.15}_{-0.02}$  &
98 eV &
8.0$\pm$0.3 &
24$\pm$1 &
--- &
0.5-25 keV: (1.19$\pm$0.01)$\times$10${^{-8}}$ &
Ref(5) \\

\emph{SUZAKU}  & 
bbody+diskbb+powerlaw & 
reflionx &
--- &
---  &
---  &
6$\pm$1 &
16$\pm$1 &
2.6$\pm$0.1 &
0.5-25 keV: (1.32$\pm$0.08)$\times$10${^{-8}}$ &
Ref(5) \\

\emph{XMM-Newton}  & 
bbody+diskbb+powerlaw & 
--- &
diskline &
6.66 - 6.97   &
38 - 50 eV &
14 - 26 &
13 - 32 &
--- &
0.5-25 keV: (0.6-0.7)$\times$10${^{-8}}$ &
Ref(5) \\

\emph{XMM-Newton}  & 
bbody+diskbb+powerlaw & 
reflionx &
--- &
--- &
---  &
15 - 107 &
3 - 9 &
2.6 - 2.8 &
0.5-25 keV: (0.6-0.7)$\times$10${^{-8}}$ &
Ref(5) \\

\hline 

\emph{NuSTAR}  & 
bbody+diskbb+powerlaw & 
--- &
kerrdisk &
6.97$\pm$0.01 &
91$\pm$2 eV &
10.6$\pm$0.6 &
18$\pm$2  &
--- &
(0.5-40 keV: 1.5$\times$10${^{-8}}$ &
Ref(6) \\

\emph{NuSTAR}  & 
bbody+diskbb+powerlaw & 
reflionx &
--- &
--- &
---  &
6 - 8.3  &
<10 &
2.30 - 2.60 &
--- &
Ref(6) \\

\hline

\emph{Chandra}  & 
bbody+diskbb+powerlaw & 
--- &
diskline &
6.97$\pm$0.02  &
149$\pm$15 eV &
7.7$\pm$0.1 &
24$\pm$1 &
--- &
--- &
Ref(7) \\

\emph{Chandra}  & 
bbody+diskbb+powerlaw & 
reflionx &
--- &
--- &
---  &
7.1$^{+1.1}_{-0.6}$ &
29$\pm$1 &
2.5$^{+0.9}_{-0.6}$  &
--- &
Ref(7) \\

\emph{Chandra}  & 
bbody+diskbb+powerlaw & 
xillver &
--- &
--- &
---  &
8.4$^{+1.1}_{-0.3}$ &
33$\pm$1 &
2.2$^{+0.7}_{-0.5}$ &
--- &
Ref(7) \\

\hline

\hline   
         
\end{tabular}

\tablefoot{Ref(1): \citet{Church.etal:01} - Ref(2):  \citet{Ooster.etal:01} - Ref(3): \citet{Bhatta.etal:07}- Ref(4): \citet{Cackett.etal:08} - Ref(5): \citet{Cackett.etal:10} - Ref(6): \citet{Miller.etal:13} - Ref(7): \citet{Chiang.etal:16}} \\
\end{sidewaystable*}



\begin{table*}
\caption{Results of the fit of NuSTAR and XMM-Newton spectra of Ser X-1 using Gaussian and Diskline models \label{tab:fit_1}}

\centering          
\scriptsize


\begin{tabular}{llccccc}     
\hline\hline       

Component &
Parameter &

gauss &
diskline&
gauss-pl &
diskline-pl &
diskline-pl-xmm\\

\hline
phabs & 
N$_{H}\,$($\times10^{22}$ cm$^{-2}$) & 
0.4 (f) &
0.4 (f) &
0.4 (f) &
0.4 (f) &
0.863$\pm$0.008  \\

bbody & 
$kT_{bb}\,$(keV) & 
0.47$\pm$0.03 &
0.54$\pm$0.06  & 
0.44$\pm$0.04 &
0.47$\pm$0.05 & 
0.47$\pm$0.02\\

R$_{BB}$&
 (km)  &  
46.1$\pm$6.3 &
34.3$\pm$7.7&
45.5$\pm$9.5 &
39.2$\pm$8.7 &
35.1$\pm$3.2 \\

bbody & 
Norm ($\times10^{-3}$) &  
22.6$\pm$2.3 &
21.8$\pm$0.8 &
16.9 $\pm$3.4 &
16.3$\pm$2.2 &
13.1$\pm$0.9\\

\hline

gaussian & 
E (keV)   & 
6.57$\pm$0.05 &
--- &
6.56$\pm$0.05 &
--- \\

gaussian & 
Sigma (keV)   & 
0.37$\pm$0.04&
--- &
0.39$\pm$0.04 & 
--- \\

gaussian & 
Norm ($\times10^{-3}$)   & 
4.03$\pm$0.35&
--- &
4.48$\pm$0.34 &
--- \\

\hline

diskline &
line E (keV)   &
---& 
6.54$\pm$0.04 &
---&
6.54$\pm$0.03&
6.48$\pm$0.06 \\

diskline & 
Betor  & 
--- &
-2.59$\pm$0.12&
---&
-2.54$\pm$0.13 &
-2.58$\pm$0.18  \\

diskline & 
R$_{in}\,$ ($R_g$) &
--- & 
18.6$\pm$4.9&
---&
19.2$\pm$4.7 &
22.0$^{+2.7}_{-5.2}$ \\

diskline & 
R$_{out}\,$ ($R_g$)  & 
--- &
2400(f) & 
---&
2400(f) &
2400(f)\\

diskline & 
Incl (deg)   & 
---&
40.1$\pm$3.6 &
---&
41.5$\pm$3.9 &
46.1$\pm$5.6 \\

diskline & 
Norm ($\times10^{-3}$)   & 
---&
4.38$\pm$0.47 &
---&
4.54$\pm$0.35 &
2.89$\pm$0.28 \\

\hline

nthComp &
Gamma   & 
2.41$\pm$0.04&
2.43$\pm$0.04 &
2.26$\pm$0.04&
2.27$\pm$0.04 &
2.10$^{+0.14}_{-0.06}$  \\

nthComp & 
$kT_{e}\,$(keV) & 
2.95$\pm$0.05 &
2.98$\pm$0.04 &
2.75$\pm$0.05 &
2.76$\pm$0.05&
2.27$\pm$0.16\\

nthComp & 
$kT_{bb}\,$(keV) & 
0.96$\pm$0.03& 
0.99$\pm$0.04 &
0.90$\pm$0.04& 
0.92$\pm$0.04 &  
0.92$\pm$0.06 ; 0.82$\pm$0.05 ; 0.88$\pm$0.06\\

nthComp & 
Norm ($\times10^{-3}$)   & 
219$\pm$11&
200$\pm$15 &
229$\pm$12&
217$\pm$18  &
160$\pm$13  \\

\hline

powerlaw  &  
Index\_pl &
--- &
--- &
3.20(f) &
3.20(f) &
3.20(f) \\

powerlaw  &  
Norm &
--- &
---&
0.84$\pm$0.12 &
0.82$\pm$0.13 &
0.72$\pm$0.04 \\

\hline

gau-rgs & 
E (keV)   & 
--- &
--- &
--- &
--- &
0.528 (f)\\

gau-rgs & 
Sigma ($\times10^{-3}$ keV)   & 
---&
--- &
--- &
--- &
2.19 (f) \\

gau-rgs & 
Norm ($\times10^{-3}$)   & 
---&
--- &
--- &
--- &
-18.4 (f) \\

\hline

gau-rgs & 
E (keV)   & 
--- &
--- &
--- &
--- &
0.541 (f)\\

gau-rgs & 
Sigma ($\times10^{-3}$ keV)   & 
---&
--- &
--- &
--- &
1.36 (f) \\

gau-rgs & 
Norm ($\times10^{-3}$)   & 
---&
--- &
--- &
--- &
57.1 (f) \\

\hline

gau-rgs & 
E (keV)   & 
--- &
--- &
--- &
--- &
0.714$\pm$0.02\\

gau-rgs & 
Sigma ($\times10^{-3}$ keV)   & 
---&
--- &
--- &
--- &
5.8$\pm$0.6 \\

gau-rgs & 
Norm ($\times10^{-3}$)   & 
---&
--- &
--- &
--- &
-12.1$\pm$0.7 \\

\hline

- &
Eq.W (eV)&
76$\pm$6&
85$\pm$7 &
84$\pm$6&
89$\pm$9 &
72$\pm$16 ; 93$\pm$18 ; 79$\pm$16 \\

- &
Obs. Flux &
5.25$\pm$0.03 &
5.27$\pm$0.03 &
5.27$\pm$0.02 &
5.27$\pm$0.02 &
3.68$\pm$0.24 \\

- &
Luminosity &
3.72$\pm$0.02 &
3.72$\pm$0.02 &
3.73$\pm$0.02 &
3.73$\pm$0.02 &
2.62$\pm$0.17 \\

\hline
 $\chi^{2}_{red}\,$(d.o.f.)   &
 - &
1.2750(915)&
1.2186(913) &
1.14134(914)&
1.0961(912) &
1.3521(4546) \\ 

\hline   
         
\end{tabular}

\tablefoot{Flux and luminosity are obtained for the 3--40 keV energy band. Fluxes  units are  
10$^{-9}$ (ergs/cm$^{2}$/sec), whereas luminosities units are 10$^{37}$ (ergs/sec)  . The seed-photon temperature was left free to vary among the three different XMM-Newton observations, this is why we report three values for this parameters in the XMM-Newton fitting results (see text for more details).
The values of the parameter Eq. W. refers to the equivalent width of the the iron line at 6.48 keV detected in each observation. Errors are reported with a 90\% confidence. 
R$_{BB}$ and luminosities are estimated assuming a distance of 7.7 kpc \citep{Galloway.etal:08}} \\
\end{table*}


\begin{table*}
\caption{Results of the fit of NuSTAR and XMM-Newton spectra of Ser X-1 using rdblur combined with rfxconv or reflionx \label{tab:fit_2}}

\centering          
\scriptsize


\begin{tabular}{llccccc}     
\hline\hline       

Component &
Parameter &

rdb-rfxconv &
rdb-reflio&
rdb-rfxconv-pl&
rdb-reflio-pl &
rdb-rfxconv-pl-xmm  \\

\hline

phabs & 
N$_{H}\,$($\times10^{22}$ cm$^{-2}$) & 
0.4 (f) &
0.4 (f) &
0.4 (f) &
0.4 (f) &
0.896$\pm$0.005 \\

bbody & 
$kT_{bb}\,$(keV) & 
0.71$\pm$0.02 &
0.80$\pm$0.02 & 
0.54$^{+0.05}_{-0.02}$ &
0.54$\pm$0.06 &
0.39$\pm$0.04 \\

R$_{BB}$ &
(km)  &  
23.6$\pm$1.3 &
15.9$\pm$0.8&
24.7$\pm$7.9 &
19.2$\pm$4.6 &
49.4$\pm$10.6\\

bbody & 
Norm ($\times10^{-3}$) &  
30.9$\pm$0.5 &
22.5 $\pm$0.6 &
11.3$^{+3.3}_{-6.1}$ &
6.8 $\pm$1.2 &
12.3 $\pm$1.6 \\

\hline

highecut &
E$_{cut}\,$(keV)   & 
--- &
0.1 (f) &
--- & 
0.1 (f) &
--- \\

highecut &
E$_{fold}\,$(keV)   & 
--- &
8.61$\pm$0.19 & 
--- &
5.04$\pm$0.09  &
--- \\

\hline

rdblur & 
Betor  & 
-3.02$\pm$0.36 &
-2.49$\pm$0.15 & 
-2.64$\pm$0.16 &
-2.53$\pm$0.14 &
-2.46$^{+0.56}_{-0.42}$ \\

rdblur & 
R$_{in}\,$ ($R_g$) &
7.7$\pm$1.3 &
15.5$\pm$4.6&
13.4$\pm$2.8&
13.2$\pm$3.1&
14.2$^{+9.5}_{-4.6}$\\

rdblur & 
R$_{out}\,$ ($R_g$)  & 
2400(f) &
2400(f) &
2400(f) &
2400(f) &
2400(f) \\

rdblur & 
Incl (deg)   & 
29.2$\pm$1.8 &
32.2$\pm$1.7 & 
27.1$\pm$1.9 &
28.8$\pm$2.4&
27(f) \\

\hline

reflionx & 
Gamma   & 
--- &
2.88$\pm$0.08 & 
--- &
1.51$\pm$0.03 &
--- \\

reflionx & 
$\xi$   & 
--- & 
4990$^{+695}_{-2350}$ &
--- &
490$^{+21}_{-98}$  &
---  \\

reflionx & 
Norm ($\times10^{-5}$)   & 
--- & 
1.97$\pm$0.59 & 
--- &
10.7$\pm$3.5 &
--- \\

\hline

rfxconv & 
rel\_refl    & 
0.55$\pm$0.04 &
--- & 
0.24$\pm$0.04 &
---  &
0.183$\pm$0.022  \\

rfxconv & 
cosIncl  & 
0.88(f)&
--- & 
0.88(f) &
--- &
0.891(f) \\

rfxconv & 
log($\xi$)   & 
2.68$\pm$0.05 &
--- & 
2.69$^{+0.02}_{-0.11}$&
--- &
3.04$\pm$0.11\\

\hline

nthComp &
Gamma   & 
3.55$\pm$0.18 &
2.88$\pm$0.08 & 
2.17$\pm$0.04 &
1.51$\pm$0.03 &
2.45$\pm$0.22 \\

nthComp & 
$kT_{e}\,$(keV) & 
4.36$^{+0.57}_{-0.23}$ &
3.19$\pm$0.08 &
2.70$\pm$0.04 &
5.05$\pm$0.09 &
3.83$^{+1.91}_{-1.02}$ \\

nthComp & 
$kT_{bb}\,$(keV) & 
1.51$\pm$0.04 &
1.43$\pm$0.05 &
0.93$\pm$0.07&
1.04$\pm$0.18 &  
0.85$\pm$0.05  ;  0.76$\pm$0.06 ;  0.82$\pm$0.06 \\

nthComp & 
Norm ($\times10^{-3}$)   & 
71.2$\pm$7.2  &
69.7$\pm$4.2 &
192$\pm$24 &
286$^{+18}_{-22}$ &
205$\pm$21 \\

\hline

powerlaw  &  
Index\_pl &
--- &
--- &
3.21$\pm$0.24  &
3.20(f) &
3.98$\pm$0.31  \\

powerlaw  &  
Norm &
--- &
--- &
1.08$^{+1.12}_{-0.72}$ &
0.82$\pm$0.13 &
0.68$\pm$0.05\\

\hline

- &
Obs. Flux &
5.26$\pm$0.15 &
5.27$\pm$0.17 &
5.27$\pm$0.62 &
5.27$\pm$0.55 &
4.12$\pm$0.38\\

- &
Luminosity &
3.73$\pm$0.11 &
3.74$\pm$0.12 &
3.74$\pm$0.44 &
3.74$\pm$0.39 &
2.93$\pm$0.27 \\

\hline
 $\chi^{2}_{red}\,$(d.o.f.)   &
 - &
1.0983(913) & 
1.0838(913) &
1.0017(911) & 
1.0123(912) & 
1.33762(4546)\\

\hline   
         
\end{tabular}

\tablefoot{For each fit, the abundance of iron in the reflection models
was kept frozen: Fe/solar = 1. 
Flux and luminosity are obtained for the 3--40 keV energy band. Fluxes  units are  
10$^{-9}$ (ergs/cm$^{2}$/sec), whereas luminosities units are 10$^{37}$ (ergs/sec)  . The seed-photon temperature was left free to vary among the three different XMM-Newton observations, this is why we report three values for this parameters in the XMM-Newton fitting results (see text for more details).  Errors are reported with a 90\% confidence. 
R$_{BB}$ and luminosities are estimated assuming a distance of 7.7 kpc \citep{Galloway.etal:08}} \\
\end{table*}


\normalsize
 

\bibliographystyle{aa} 
\bibliography{serx1}
\begin{appendix}
\section{Models including \texttt{kerrconv}}
\label{sec:appen}
From the spectral analysis described in Sec. \ref{sec:spec}, we find that our 
best fit obtained using \texttt{rdblur} as smearing component gives a 
soft blackbody temperature of 0.54$\pm$0.06 keV and a radius of the emitting 
region of $25 \pm 6$ km, a temperature of the seed photons for the Comptonization 
of 0.93$\pm$0.07 keV, an electron temperature of the Comptonizing corona of  
2.70$\pm$0.04 keV and a photon index of the primary Comptonized component of  
2.17$\pm$0.04,  whereas the photon index of the hard power-law tail is steeper, 
around 3.2.
The reflection component gives a reflection amplitude of 0.24$\pm$0.04 and a 
ionization parameter of log($\xi$) = 2.69$^{+0.02}_{-0.11}$. Finally, the 
smearing of the reflection component gives an inner disk radius of R$_{in}$ = 
13.4$\pm$ 2.8R$_{g}$, compatible with the radius inferred from the blackbody
component, and an emissivity index of the disk equal to -2.64$\pm$0.16, 
whereas the inclination angle of the disk with respect to the line of sight
results equal to 27.1$\pm$1.9$^\circ$.
The analysis of \emph{XMM-Newton} spectra, carried out using the same models 
adopted to fit the \emph{NuSTAR} spectra, gave values of the parameters
compatible to those described above, although the two observations are
not simultaneous. In particular in this case we find R$_{in}$ 
14.2$^{+9.5}_{-4.6}$ R$_g$, a reflection amplitude of 0.183$\pm$0.003 and an 
ionization parameter of log($\xi$) = 3.04$\pm$0.11, a temperature of the seed photons in 
the range $0.76-0.85$ keV, a photon index of the primary Comptonized component of 
2.45$\pm$0.22 keV. In other words, the \emph{XMM-Newton} spectra independently
confirm the results obtained for the \emph{NuSTAR} spectra.

In order to check the presence of a non-null spin parameter of the neutron 
star, we fitted the \emph{NuSTAR} spectra using reflection components 
convolved with \texttt{kerrconv} instead 
of \texttt{rdblur}. Kerrconv convolves the spectrum with the smearing 
produced by a kerr disk model. It features the dimensionless 'a' parameter 
that characterize the spin of the system. 
We have performed our fit first leaving 'a' as a free parameter and then
fixing it to the following three values, 0, 0.12, 0.14. The model with reflionx and 
'a' treated as free parameter is called \emph{ker-reflio-af}, whereas for 
$a=0$, $a=0.12$ and $a=0.14$ the models are called \emph{ker-reflio-a0}, 
\emph{ker-reflio-a012}, and \emph{ker-reflio-a014}, respectively. 
In the same way, the model with rfxconv and 'a' treated as free parameter 
is called \emph{ker-rfxconv-af}, whereas for $a=0$, $a=0.12$ and $a=0.14$ 
the models are called \emph{ker-rfxconv-a0}, \emph{ker-rfxconv-a012}, and 
\emph{ker-rfxconv-a014}, respectively. All the models fit the data well;
reduced $\chi^{2}\,$ are between 1.08 and 1.18 and residuals are basically 
identical . Moreover the best-fit 
values of all parameters are very similar to the case with $a=0$ and to the 
values we get using \texttt{rdblur} instead of \texttt{kerrconv}. The fit is 
therefore insensitive to the spin parameter, although there is 
a slight preference of the fit towards low values ($a<0.04$).
It is worth noting that in all best fit residuals a feature is present 
at about 3.9 keV that could be the resonance line of Ca XIX (3.9 keV). 
Moreover, again we observe high energy residuals (above 30 keV) indicating
the presence of a hard power-law component. Also in this case, we get a 
very large ionization parameter using reflionx. 

To avoid this problem, we therefore added a power-law component to the model 
obtained by the convolution of \texttt{kerrconv} with the two different reflection 
components (reflionx or rfxconv). we considered 'a' free to vary or fixed it to
three different values (0, 0.12, 0.14). In all the cases the fits are quite
good with values of the reduced $\chi^2$ from 1.0 to 1.01. Again the addition 
of the power-law proved to be highly statistically significant. The F-test 
probability of chance improvement for the addition of two parameters is,
for instance, $7.5 \times 10^{-15}$ and $9 \times 10^{-33}$ for the addition 
of a power-law to the model \emph{ker-reflio-af} and \emph{ker-rfxconv-af},
respectively.
As before, the fit is quite insensitive to the value assumed by the spin 
parameter 'a'. 
Values of the best-fit parameters are listed in Tab \ref{tab:fit_3} and \ref{tab:fit_4}.


\begin{sidewaystable*}

\caption{Results of the fit of the NuSTAR spectra using kerrconv combined with 
rfxconv or reflionx components \label{tab:fit_3}}

\centering  
        
\scriptsize



\begin{tabular}{llcccccccc}     
\hline\hline       

Component &
Parameter &

ker-reflio-af &
ker-reflio-a0 &
ker-reflio-a012 &
ker-reflio-a014 &
ker-rfxconv-af &
ker-rfxconv-a0 &
ker-rfxconv-a012 &
ker-rfxconv-a014 \\

\hline

bbody & 
$kT_{bb}\,$(keV) & 
0.79$\pm$0.02 &
0.80$\pm$0.02 &
0.80$\pm$0.03 &
0.80$\pm$0.03 &
0.70$\pm$0.02 &
0.71$\pm$0.02 &
0.71$\pm$0.03 &
0.67$\pm$0.04 \\

bbody & 
Norm ($\times10^{-3}$) &  
22.4$\pm$0.3 &
22.5$\pm$0.4 &
22.5$\pm$0.7 &
22.4$\pm$0.8 &
30.1$\pm$0.11 &
31.2$\pm$0.3 &
30.8$\pm$0.2 &
29.9$\pm$0.5 \\

\hline

highecut &
E$_{cut}\,$(keV)   & 
0.1 (f)&
0.1 (f) &
0.1 (f) &
0.1 (f) &
---&
---&
---&
--- \\

highecut &
E$_{fold}\,$(keV)   & 
8.54$\pm$0.06 &
8.56$\pm$0.13 &
8.52$\pm$0.29 &
8.59$\pm$0.21 &
---&
---&
---&
--- \\

\hline

kerrconv & 
Index   & 
2.38$\pm$0.25 &
2.53$\pm$0.16&
2.49$\pm$0.19 &
2.46$\pm$0.18 &
3.65$\pm$0.27 &
3.35$\pm$0.48 &
2.7$\pm$0.25&
6.5$\pm$1.6 \\



kerrconv & 
a   & 
0.019$^{+0.019}_{-0.021}$ &
0.0 (f) &
0.12 (f) &
0.14(f) &
0.036$\pm$0.008  &
0.0(f) &
0.12(f) &
0.14(f)\\

kerrconv & 
Incl (deg)   & 
32.2$\pm$1.9 &
32.1$\pm$1.3 &
31.9$\pm$1.6 &
32.1$\pm$2.1 &
30.4$\pm$0.4 &
30.4$\pm$1.6 &
29.4$\pm$0.4&
35.7$\pm$1.6 \\

kerrconv & 
R$_{in}\,$ ($R_g$) &
14.5$\pm$1.8 & 
18.1$\pm$5.7&
16.3$\pm$5.4&
15.7(f)&
7.8$\pm$0.4&
7.8$\pm$1.6 &
$ < 12.5$ &
6.8$\pm$0.3\\

kerrconv & 
R$_{out}\,$ ($R_g$)  & 
2400(f) & 
2400(f) &
2400(f) &
2400(f) &
2400(f) &
2400(f) &
2400(f) &
2400(f)\\

\hline

reflionx & 
Gamma   & 
2.85$\pm$0.05 &
2.86$\pm$0.05 &
2.85$\pm$0.12 &
2.87$\pm$0.08&
--- &
--- &
--- &
--- \\

reflionx & 
$\xi$   & 
3722$\pm$61&
3784$^{+2050}_{-1100}$ &
3580$^{+2140}_{-1430}$ &
4980$^{+990}_{-2270}$ &
--- &
--- &
--- &
--- \\

reflionx & 
Norm ($\times10^{-5}$)   & 
2.38$\pm$0.55 &
2.36$\pm$0.95 &
2.47$\pm$1.25 &
1.91$^{+1.62}_{-0.58}$ &
--- &
--- &
--- &
--- \\

\hline

rfxconv & 
rel\_refl    & 
--- &
--- &
--- &
--- &
0.58$\pm$0.06 &
0.58$\pm$0.03 &
0.54$\pm$0.02 &
0.69$\pm$0.03 \\

rfxconv & 
cosIncl  & 
---&
--- &
--- &
--- &
0.88(f) &
0.88(f) &
0.88(f) &
0.88(f) \\

rfxconv & 
log($\xi$)   & 
--- &
--- &
--- &
--- &
2.71$\pm$0.03 &
2.68$\pm$0.04 &
2.68$\pm$0.03 &
2.69$\pm$0.04\\

\hline

nthComp &
Gamma   & 
2.85$\pm$0.05 &
2.86$\pm$0.05 &
2.85$\pm$0.12 &
2.87$\pm$0.08 &
3.74$\pm$0.02 &
3.75$\pm$0.07 &
3.69$\pm$0.06 &
3.76$\pm$0.12\\

nthComp & 
$kT_{e}\,$(keV) & 
3.16$\pm$0.03 &
3.17$\pm$0.05 &
3.16$\pm$0.11 &
3.19$^{+0.09}_{-0.05}$ &
4.53$\pm$0.06 &
4.51$\pm$0.25 &
4.40$\pm$0.16 &
4.62$^{+0.35}_{-0.24}$ \\

nthComp & 
$kT_{bb}\,$(keV) & 
1.42$\pm$0.03 &
1.43$\pm$0.02 &
1.43$\pm$0.03 &
1.43$\pm$0.04 &
1.53$\pm$0.03 &
1.53$\pm$0.03 &
1.52$\pm$0.05 & 
1.55$\pm$0.03\\

nthComp & 
Norm ($\times10^{-3}$)   & 
70.9$\pm$1.9 &
70.3$\pm$3.5&
69.9$^{+6.6}_{-2.2}$ &
70.2$^{+4.5}_{-3.3}$ &
69.7$^{+2.9}_{-12.2}$ &
69.6$^{+1.1}_{-5.5}$ &
71.1$^{+5.9}_{-3.3}$ &
71.3$^{+1.8}_{-5.2}$ \\

\hline

- &
R$_{BB}$ (km)  &  
14.9$\pm$0.9 &
14.5$\pm$0.8&
14.6$\pm$1.1&
14.5$\pm$1.5&
21.1$\pm$1.3&
21.8$\pm$1.2 &
21.6$\pm$1.7 &
21.9$\pm$2.9 \\


\hline

 $\chi^{2}_{red}\,$(d.o.f.)   &
 - &
1.0876(912) &
1.0876(913) &
1.0859(913) & 
1.0835(914) & 
1.1797(914) &
1.0981(913) &
1.1111(913) &
1.0849(913) \\
 
\hline   
         
\end{tabular}

\tablefoot{For each fit, the following parameters were kept frozen: 
N$_{H}\,$ = 0.4 $\times10^{22}$ cm$^{-2}$, Fe/solar = 1 and 
r$_{br}$ = 6 $R_g$. The parameter r$_{br}$ in the
kerrconv model is break radius separating the inner and outer portions 
of the disk, having emissivity index Index1 and Index2, respectively, 
which in our fit are constrained to assume the same value, Index.
Errors are reported with a 90\% confidence. 
R$_{BB}$ are estimated assuming a distance of 7.7 kpc \citep{Galloway.etal:08}} \\
\end{sidewaystable*}


\begin{sidewaystable*}


 \caption{Fitting results adding a power-law to the models of Table  \label{tab:fit_4}}             
    
\centering 
         
\scriptsize



\begin{tabular}{llcccccccc}     
\hline\hline       

Component &
Parameter &

ker-reflio-af-pl &
ker-reflio-a0-pl &
ker-reflio-a012-pl &
ker-reflio-a014-pl &
ker-rfxconv-af-pl &
ker-rfxconv-a0-pl &
ker-rfxconv-a012-pl &
ker-rfxconv-a014-pl \\

\hline

bbody & 
$kT_{bb}\,$(keV) & 
0.54$\pm$0.18 & 
0.52$\pm$0.12 &
0.56$\pm$0.08 &
0.53$\pm$0.13 &
0.55$\pm$0.03 &
0.59$\pm$0.08 &
0.55$\pm$0.03 &
0.50$\pm$0.07 \\

bbody & 
Norm ($\times10^{-3}$) &  
6.8$\pm$0.3 &
8.4$\pm$1.1 &
6.7$\pm$0.7 &
8.4$\pm$0.8 &
11.9$\pm$1.6 &
14.1$\pm$0.8 &
11.9$\pm$0.4 &
10.8$\pm$1.2 \\

\hline

highecut &
E$_{cut}\,$(keV)   & 
0.1 (f) &
0.1 (f) &
0.1 (f) &
0.1 (f) &
---&
---&
---&
--- \\

highecut &
E$_{fold}\,$(keV)   & 
5.05$\pm$0.08 &
5.08$\pm$0.07 &
5.04$\pm$0.05 &
5.08$\pm$0.08 &
---&
---&
---&
--- \\

\hline

kerrconv & 
Index   & 
2.59$\pm$0.14 &
2.51$\pm$0.12&
2.54$\pm$0.08 &
2.51$\pm$0.19 &
2.72$\pm$0.27 &
2.78$\pm$0.12 &
2.71$\pm$0.13&
2.66$\pm$0.18 \\


kerrconv & 
a   & 
$ < 0.019$ &
0.0 (f) &
0.12 (f) &
0.14(f) &
0.06$^{+0.67}_{-0.02}$  &
0.0(f) &
0.12(f) &
0.14(f)\\

kerrconv & 
Incl (deg)   & 
28.3$\pm$1.7 &
28.4$\pm$0.9 &
28.3$\pm$0.5 &
28.3$\pm$1.4 &
26.0$\pm$0.9 &
26.1$\pm$0.9 &
26.1$\pm$0.8&
26.1$\pm$0.8 \\

kerrconv & 
R$_{in}\,$ ($R_g$) &
13.6$\pm$4.8 & 
12.6$\pm$1.5&
14.5$\pm$0.4&
13.8$\pm$4.5&
15.3$\pm$3.9&
13.2$\pm$3.1 &
15.6$\pm$3.8 &
15.9$\pm$3.6\\

kerrconv & 
R$_{out}\,$ ($R_g$)  & 
2400(f) & 
2400(f) &
2400(f) &
2400(f) &
2400(f) &
2400(f) &
2400(f) &
2400(f)\\

\hline

reflionx & 
Gamma   & 
1.51$\pm$0.04 &
1.52$\pm$0.03 &
1.50$\pm$0.04 &
1.52$\pm$0.04&
--- &
--- &
--- &
--- \\

reflionx & 
$\xi$   & 
497$^{+23}_{-81}$ &
496$\pm$17 &
501$\pm$19 &
497$^{+14}_{-79}$ &
--- &
--- &
--- &
--- \\

reflionx & 
Norm ($\times10^{-5}$)   & 
10.5$\pm$1.8 &
9.9$\pm$0.8 &
10.4$\pm$3.2 &
9.9$^{+2.8}_{-1.1}$ &
--- &
--- &
--- &
--- \\

\hline

rfxconv & 
rel\_refl    & 
--- &
--- &
--- &
--- &
0.24$\pm$0.04 &
0.27$\pm$0.03 &
0.25$\pm$0.03 &
0.24$\pm$0.03 \\

rfxconv & 
cosIncl  & 
---&
--- &
--- &
--- &
0.88(f) &
0.88(f) &
0.88(f) &
0.88(f) \\

rfxconv & 
log($\xi$)   & 
--- &
--- &
--- &
--- &
2.71$\pm$0.03 &
2.69$\pm$0.04 &
2.71$\pm$0.05 &
2.69$\pm$0.05\\

\hline

nthComp &
Gamma   & 
1.51$\pm$0.04 & 
1.52$\pm$0.03  &
1.50$\pm$0.04 &
1.52$\pm$0.04 &
2.19$\pm$0.04 &
2.24$\pm$0.05 &
2.19$\pm$0.05 &
2.15$\pm$0.04\\

nthComp & 
$kT_{e}\,$(keV) & 
5.04$\pm$0.08 &
5.08$\pm$0.07 &
5.04$\pm$0.05 &
5.08$\pm$0.08 &
2.71$\pm$0.05 &
2.76$\pm$0.08 &
2.71$^{+0.02}_{-0.06}$ &
2.68$^{+0.07}_{-0.03}$ \\

nthComp & 
$kT_{bb}\,$(keV) & 
1.04$\pm$0.22 &
1.05$\pm$0.06 &
1.04$\pm$0.04 &
1.04$\pm$0.14 &
0.94$\pm$0.07 &
1.01$\pm$0.16 &
0.94$\pm$0.05 & 
0.90$\pm$0.09 \\

nthComp & 
Norm ($\times10^{-3}$)   & 
287$\pm$19 &
289$\pm$77&
501$\pm$15 &
289$\pm$38 &
187$^{+5}_{-11}$ &
161$^{+46}_{-5}$ &
187$^{+9}_{-18}$ &
210$^{+7}_{-38}$ \\

\hline

powerlaw  &  
Index\_pl &
3.20$^{+0.33}_{-1.02}$ &
3.20(f) &
3.20(f) &
3.09$^{+0.57}_{-1.02}$ & 
3.20(f)&
3.20(f)&
3.20(f) &
3.20(f) \\

powerlaw  &  
Norm &
0.81$\pm$0.42 &
0.82$\pm$0.12 &
0.80$\pm$0.08 &
0.54$^{+2.13}_{-0.34}$ &
1.04$\pm$0.09 & 
0.98$\pm$0.08 &
1.04$\pm$0.13 &
1.06$\pm$0.11 \\

\hline
 
 - &
R$_{BB}$ (km)  &  
17.6$\pm$10.8 &
20.9$\pm$8.9&
16.2$\pm$5.3&
20.3$\pm$8.9&
22.4$\pm$2.9&
21.2$\pm$5.1 &
22.4$\pm$2.4 &
25.8$\pm$7.4 \\


\hline

 $\chi^{2}_{red}\,$(d.o.f.)   &
 - &
1.0148(910) f&
1.0123(912) &
1.0083(912) & 
1.0138(911) & 
1.0016(911) &
1.0023(912) &
1.0008(912) &
1.0006(912) \\
 
\hline   
         
\end{tabular}

\tablefoot{For each fit, the following parameters were kept frozen: 
N$_{H}\,$ = 0.4 $\times10^{22}$ cm$^{-2}$, Fe/solar = 1 and 
r$_{br}$ = 6 $R_g$. The parameter r$_{br}$ in the
kerrconv model is break radius separating the inner and outer portions 
of the disk, having emissivity index Index1 and Index2, respectively, 
which in our fit are constrained to assume the same value, Index.
Errors are reported with a 90\% confidence. 
R$_{BB}$ are estimated assuming a distance of 7.7 kpc \citep{Galloway.etal:08}} \\

\end{sidewaystable*}

\end{appendix}

\normalsize

\end{document}